\documentclass[reprint,superscriptaddress,amsmath,amsthm,amssymb,aps,prb]{revtex4-1}
\usepackage{color}
\usepackage{graphicx}
\usepackage{dcolumn}
\usepackage{bm}
\usepackage[colorlinks=true 
,urlcolor=blue
,anchorcolor=bluerisk 
,citecolor=red
,filecolor=blue
,linkcolor=blue
,menucolor=blue
,pagecolor=blue
,linktocpage=true
,pdfproducer=medialab
]{hyperref}
\usepackage{ragged2e}
\usepackage{graphicx}
\usepackage{caption}
\usepackage[labelformat=simple]{subcaption}

\usepackage{feynmp}  \unitlength = .075mm 
\usepackage{comment}

\usepackage{soul} 
		  \setstcolor{red}   
		  \setul{1ex}{0.2ex} 
\usepackage{cancel}

\begin{document}

\preprint{APS}
\title{Renormalization group analysis on a neck-narrowing Lifshitz transition in the presence of weak short-range interactions in two dimensions}

\author{Sedigh Ghamari}
\affiliation{Department of Physics \& Astronomy, McMaster University, 1280 Main St. W., Hamilton Ontario L8S 4M1, Canada.}

\author{Sung-Sik Lee} 
\affiliation{Department of Physics \& Astronomy, McMaster University, 1280 Main St. W., Hamilton Ontario L8S 4M1, Canada.}
\affiliation{Perimeter Institute for Theoretical Physics, 31 Caroline St. N., Waterloo Ontario N2L 2Y5, Canada.}

\author{Catherine Kallin}
\affiliation{Department of Physics \& Astronomy, McMaster University, 1280 Main St. W., Hamilton Ontario L8S 4M1, Canada.}
\affiliation{The Canadian Institute for Advanced Research, Toronto, Ontario M5G 1Z8, Canada.}
\affiliation{Kavli Institute for Theoretical Physics, University of California, Santa Barbara, California 93106-9530, USA.}

\date{\textcolor{blue}{\textit{\today}}}

\begin{abstract}
We study a system of weakly interacting electrons described by the energy dispersion $\xi(\mathbf{k}) = k_x^2 - k_y^2 - \mu$ in two dimensions within a renormalization group approach. This energy dispersion exhibits a neck-narrowing Lifshitz transition at the critical chemical potential $\mu_c=0$ where a van Hove singularity develops. Implementing a systematic renormalization group analysis of this system has long been hampered by the appearance of nonlocal terms in the Wilsonian effective action. We demonstrate that nonlocality at the critical point is intrinsic, and the locality of the effective action can be maintained only away from the critical point. We also point out that it is crucial to introduce a large momentum cutoff to keep locality even away from the critical point. Based on a local renormalization group scheme employed near the critical point, we show that, as the energy scale $E$ is lowered, an attractive four-fermion interaction grows as $\log^2 E$ for $E > \mu$, whereas it retains the usual BCS growth, $-\log E$, for $E < \mu$. Starting away from the critical point, this fast growth of the pairing interaction suggests that the system becomes unstable toward a superconducting state well before the critical point is reached.
\end{abstract}
%
%
%
%
%
\begin{fmffile}{Feynman_Graphs}  
\maketitle
%
%
%

\section{Introduction}
\label{introduction}

\begin{figure}[h] 
        \centering
        \begin{subfigure}[b]{0.15\textwidth}
                \includegraphics[scale=0.2]{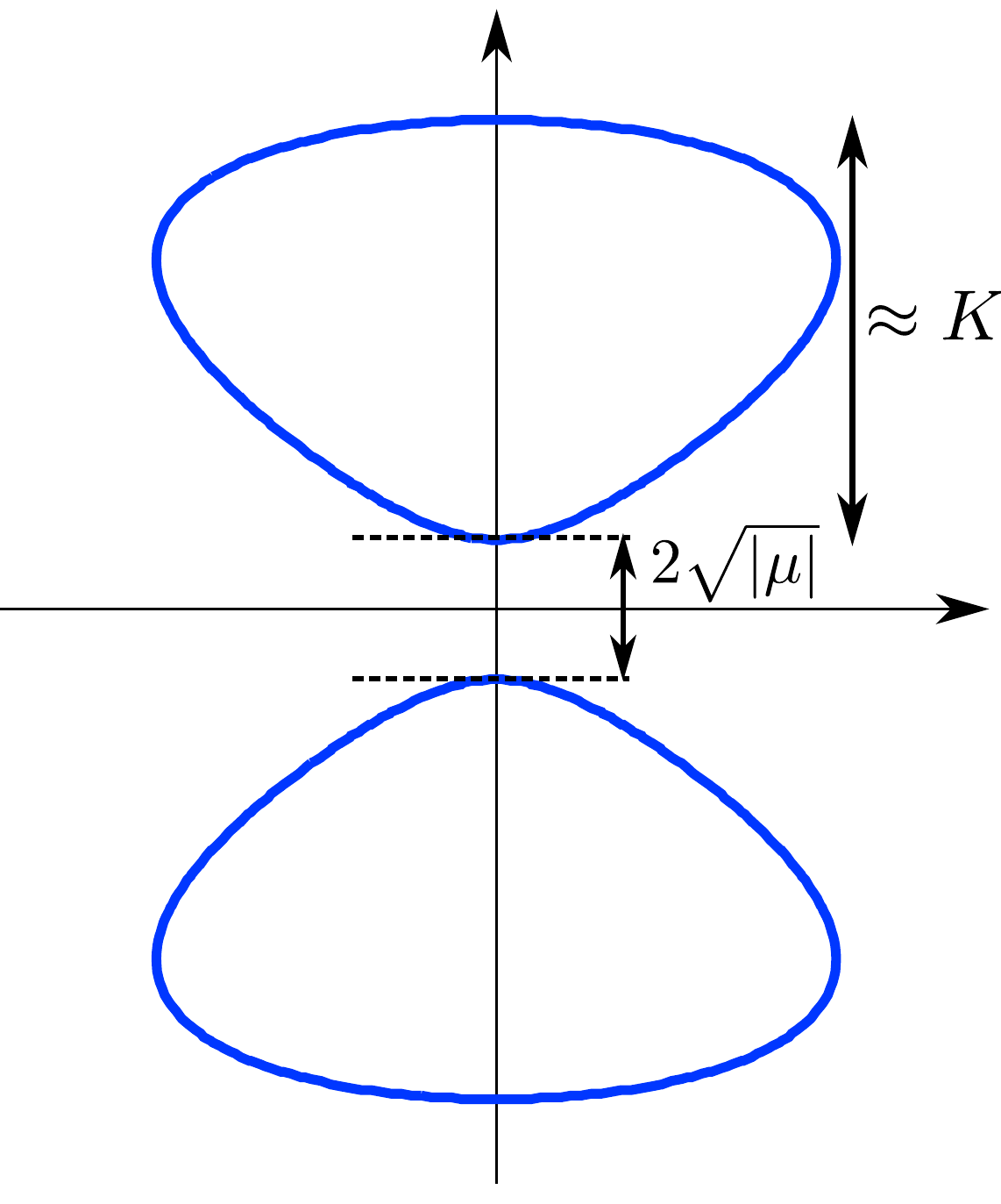}
	
		\vspace{.25cm}
		\includegraphics[width=1\textwidth]{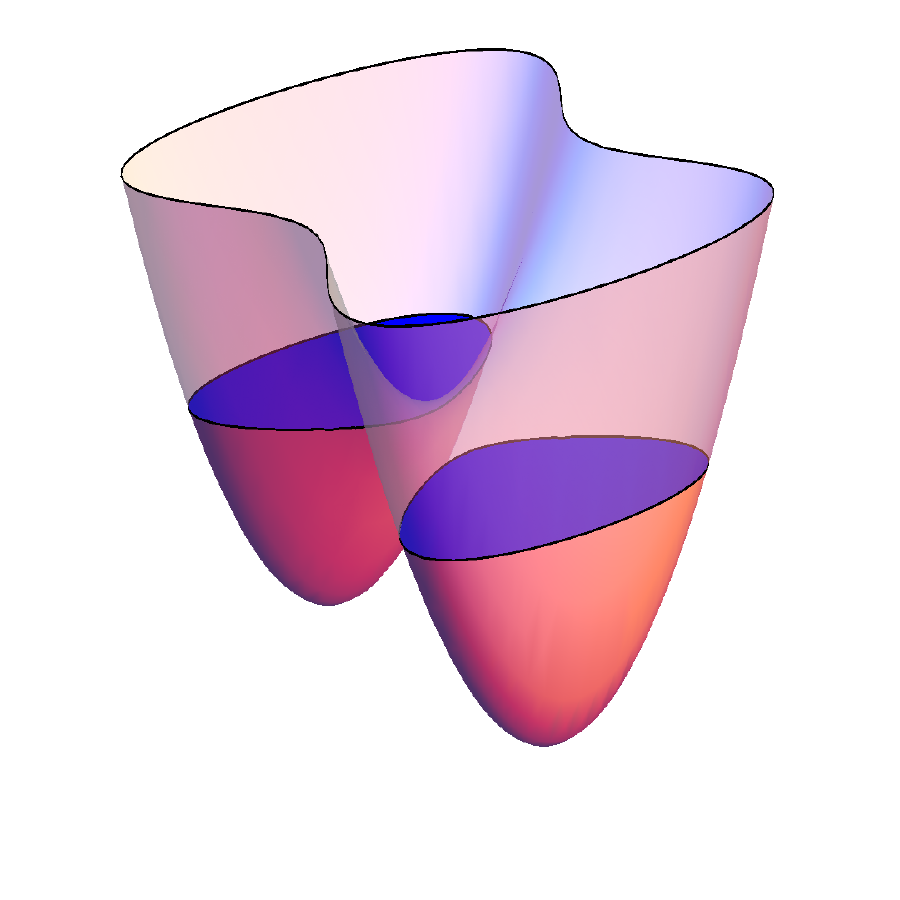}
                \caption{$\mu<0$}
                \label{Cartoon_Lifshitz_neck-narrowing:muL0}
        \end{subfigure}
	~
        \begin{subfigure}[b]{0.15\textwidth}
                \includegraphics[scale=0.2]{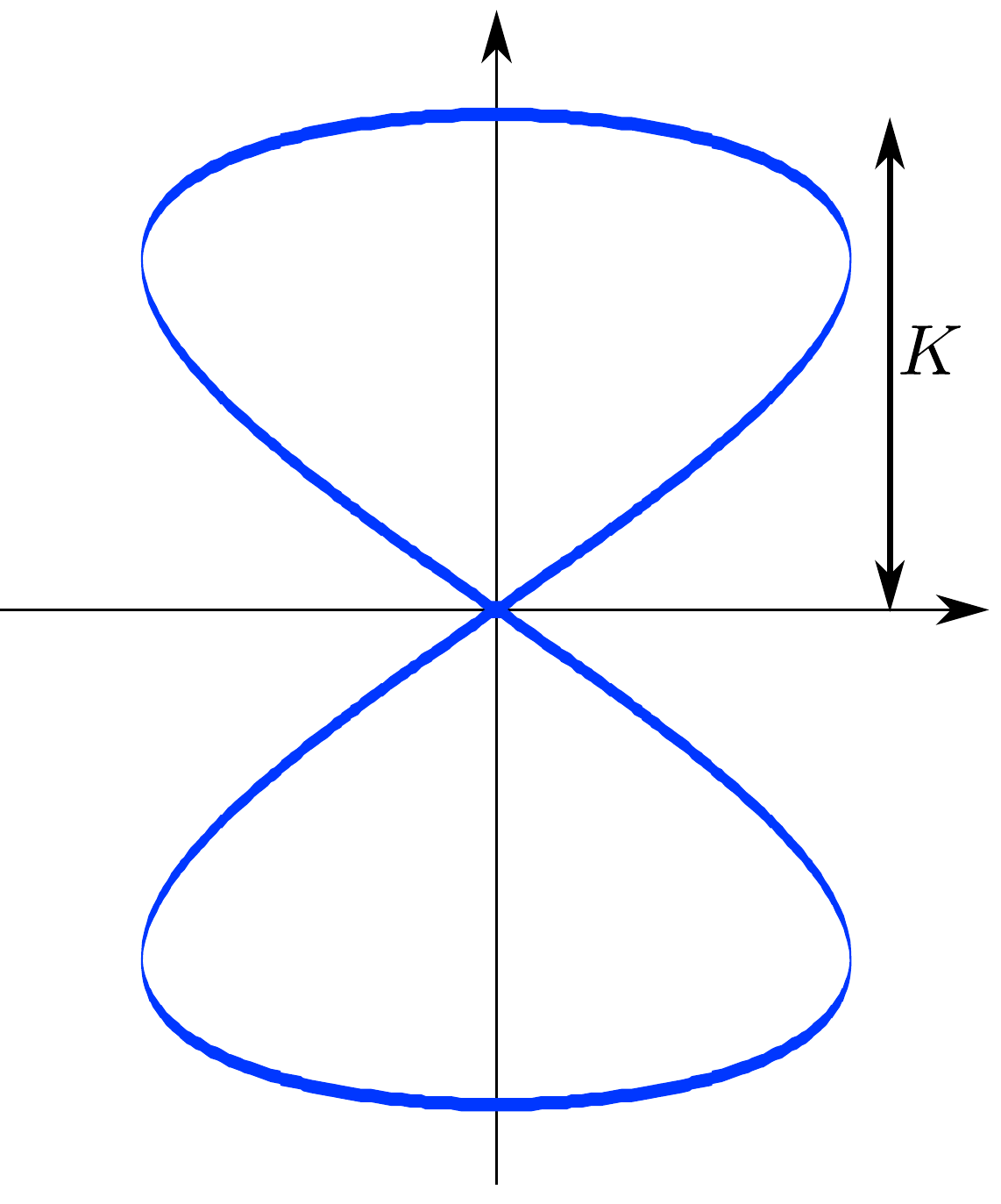}
	
		\vspace{.25cm}
		\includegraphics[width=1\textwidth]{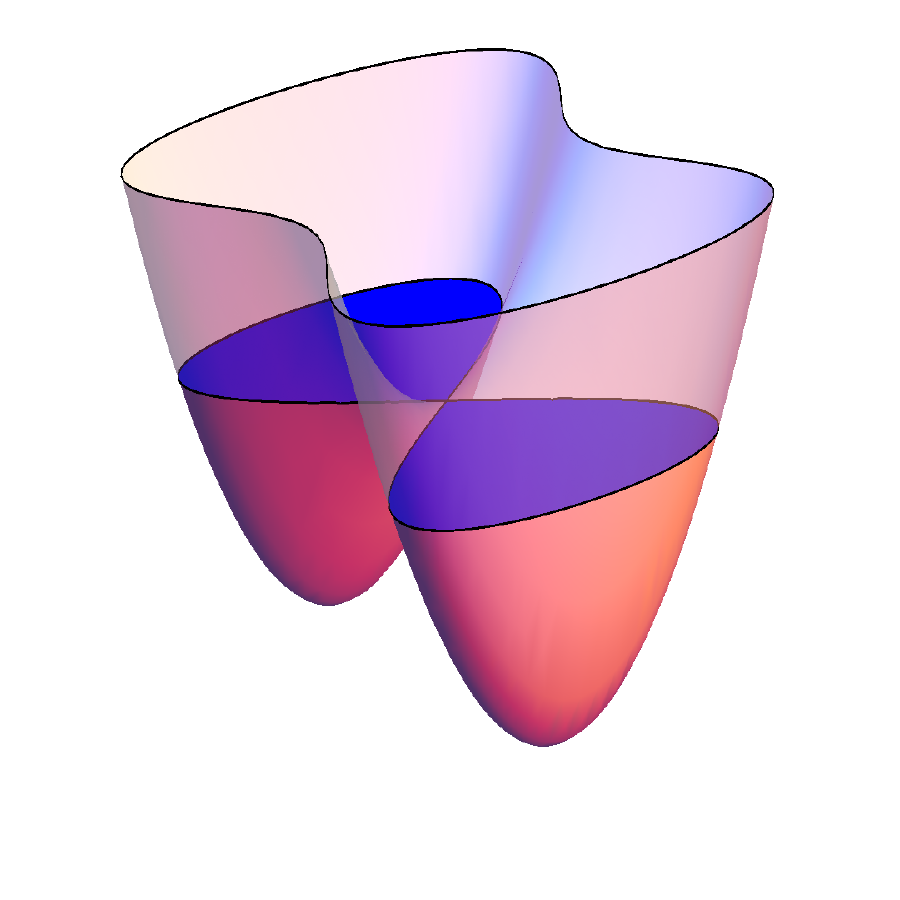}
                \caption{$\mu=0$}
                \label{Cartoon_Lifshitz_neck-narrowing:mu0}
        \end{subfigure}
	~
        \begin{subfigure}[b]{0.15\textwidth}
                \includegraphics[scale=0.2]{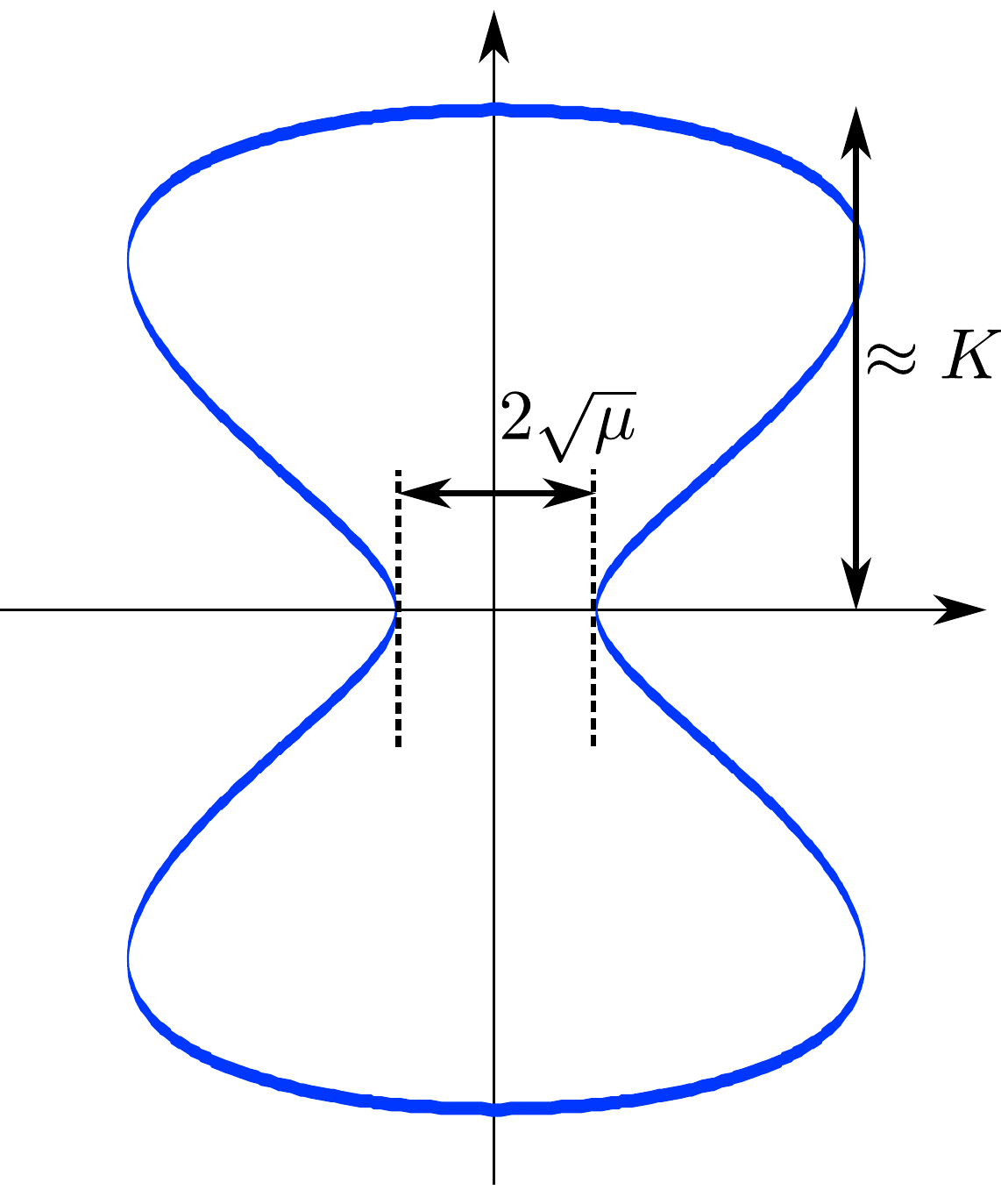}
		
		\vspace{.25cm}
		\includegraphics[width=1\textwidth]{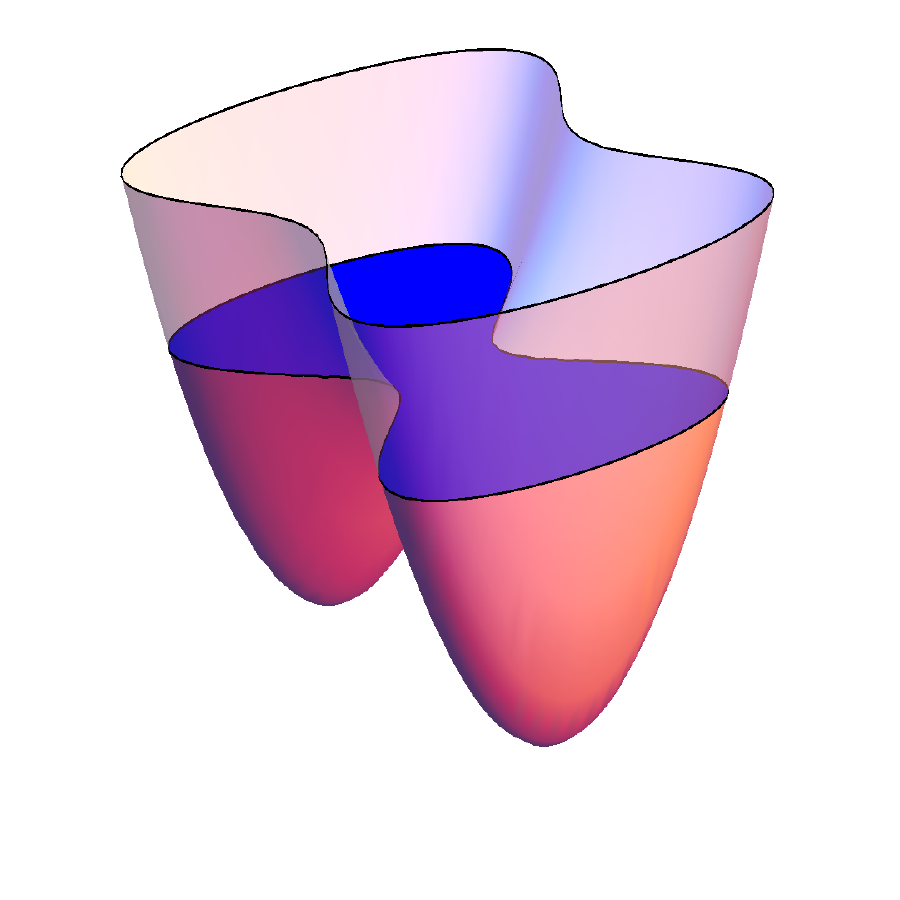}
                \caption{$\mu>0$}
                \label{Cartoon_Lifshitz_neck-narrowing:muG0}
        \end{subfigure}

        \caption[]{\justifying A neck-narrowing Lifshitz transition in two dimensions for a noninteracting Fermion model with the dispersion $\varepsilon(\mathbf{k})=k_x^2-k_y^2+\frac{1}{K^2}k_y^4$ at (a) $\mu<0$, where the Fermi surface is made up of two separate lobes, (b) the critical point, $\mu_c=0$, where the two lobes first touch and a van Hove singularity is developed, and (c) $\mu>0$ at which point a smooth monolithic Fermi surface is formed.}
	\label{Cartoon_Lifshitz_neck-narrowing}
\end{figure}

Lifshitz transitions occur when two parts of a Fermi surface collide (neck-narrowing Lifshitz transition) or a new Fermi pocket appears/disappears in momentum space (pocket-disappearing Lifshitz transition) as some parameter, such as chemical potential, is tuned.~\cite{Lifshitz1960,Yamaji2006} Such transitions are topological in nature without involving any change in symmetry.~\cite{WEN2004,Rodney2013} Near a neck-narrowing Lifshitz transition in two dimensions (see Fig.~[\ref{Cartoon_Lifshitz_neck-narrowing}]), the density of states (DOS) is proportional to $\log\frac{K}{\sqrt{\smash[b]{|\mu|}}}$, where $K$ is the size of the Fermi surface and $\mu$ is the chemical potential relative to the critical point. At the critical point, the DOS is logarithmically divergent, indicating a van Hove singularity.

The fate of van Hove singularities in the presence of short-range interactions in two dimensions has been the focus of many  studies.~\cite{Schulz1987, Lederer1987, Gonzalez1997, Alvarez1998, Furukawa1998, Furukawa2000, Gonzalez2001, LeHur2009, Nandkishore2012, Kapustin2012, Gonzalez2013, Yudin2013, Khavkine2004, Yamase2005, Zanchi1996, Zanchi2000, Reiss2007, Honerkamp2002, Neumayr2003, Raghu2010} However, a systematic renormalization group study is still lacking as usual perturbative local renormalization group (RG) schemes cannot be applied when nonlocal terms are generated in the effective action, as in this system. The nonlocal nature of the system can be inferred from the one-loop quantum effective action in the particle-particle channel [see Fig.~\ref{1loop_interaction:PP}], which is proportional to $\log^2 (\Lambda/E)$,~\cite{Furukawa1998,Nandkishore2012,Gonzalez1997,Kapustin2012} where $\Lambda$ is a UV energy cutoff and $E$ is an external energy. Here, one of the logarithms arises from the usual loop corrections to the coupling that is marginal in two dimensions, and the other originates from the divergent DOS.~\cite{Gonzalez2000} This log-squared term in the quantum effective action gives rise to a four fermion vertex proportional to $-2\log(E/\Lambda)$ in the Wilsonian effective action. Note that nonanalyticity in the energy-momentum space translates into nonlocality of the action in real space.

The nonlocality of the Wilsonian effective action poses a serious problem to the implementation of a systematic RG approach. Once a nonlocal term appears in the Wilsonian effective action, infinitely many other nonlocal terms can subsequently get generated as high-energy modes are further integrated out. The proliferation of nonlocal terms makes it impossible to constrain the form of the effective action to a finite set of couplings based on a gradient expansion. For instance, the appearance of a nonlocal density-density interaction vertex of the form $\log (E/\Lambda)$ can, at later stages of RG, give rise to new nonlocal vertices such as $\log^n(E/\Lambda)$ (with $n > 1$), resulting in a cascade of nonlocal terms. If the form of nonlocal terms are constrained by some symmetry, the proliferation of nonlocal terms can, in principle, be contained. Nevertheless, this is not the case in the problem at hand. In some of the previous studies,~\cite{Furukawa1998,Nandkishore2012}, $\beta$ functions were defined in terms of derivatives of the quantum effective action with respect to $\log^2 \Lambda$, however, this amounts to ignoring nonlocal terms.

\begin{figure}[h] 
        \centering
        \begin{subfigure}[m]{0.225\textwidth}
        	\includegraphics[width=1.\textwidth]{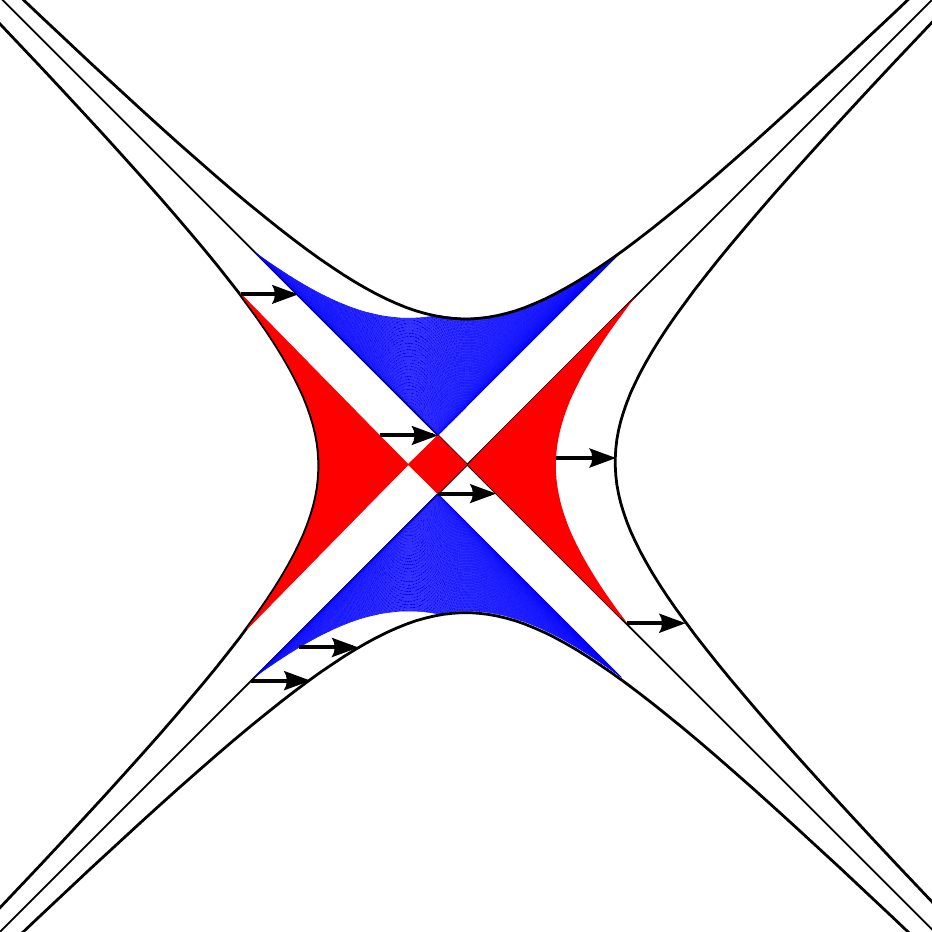}
                \caption{}
                \label{PP_Phase_Space:I}
        \end{subfigure}
	~
        \begin{subfigure}[m]{0.225\textwidth}
		\includegraphics[width=1.\textwidth]{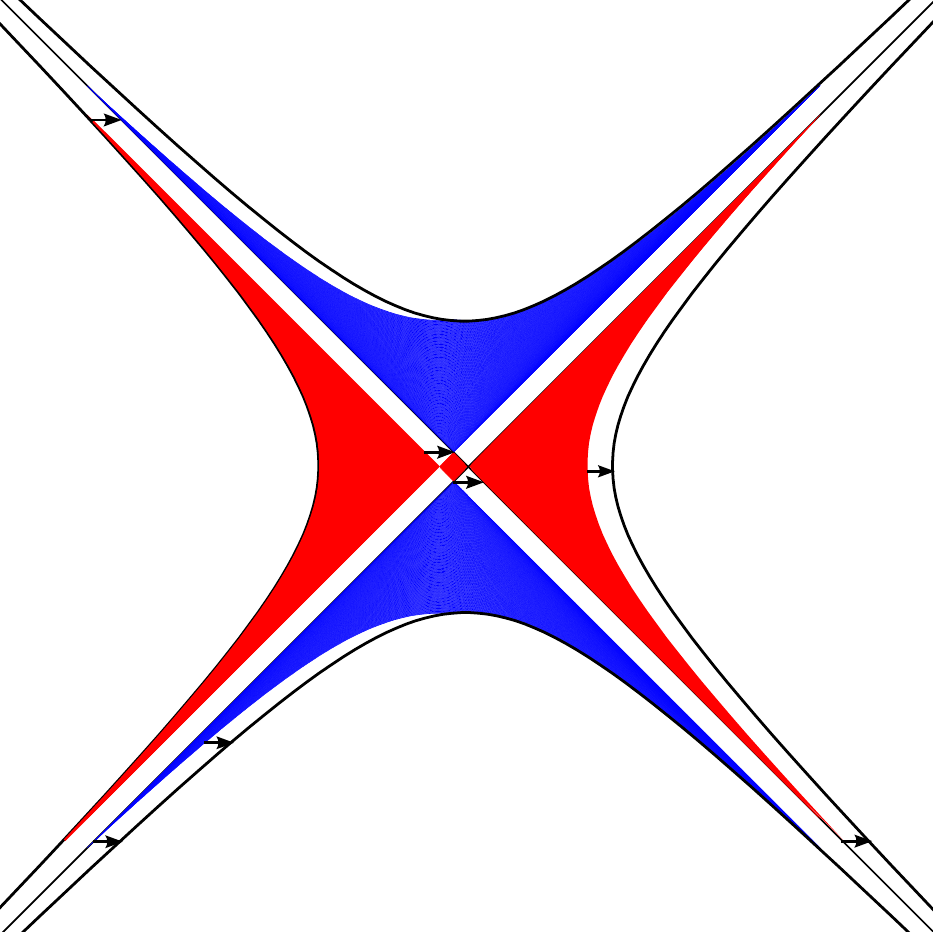}
                \caption{}
                \label{PP_Phase_Space:II}
        \end{subfigure}
	
        \caption[]{\justifying The (red and blue) shaded regions indicate the phase space available for the intermediate states in the one-loop particle-particle diagram $\Gamma_{\text{\tiny PP}}(\mathbf{q}=q\hat{x})$ [see Fig.~\ref{1loop_interaction:PP} and Eq.~(\ref{1-loop_PP})] for the dispersion relation $\epsilon(\mathbf{k}) = k_x^2-k_y^2$. The external momentum $\mathbf{q}$, denoted by arrows in the figures, in (a) is twice of that in (b). As $|\mathbf{q}|$ decreases a larger region contributes to the one-loop particle-particle diagram, resulting in a singular Wilsonian effective action in the small $|\mathbf{q}|$ limit. Here, a sharp energy cutoff $\Lambda=25$ is imposed. 
	}
	\label{PP_Phase_Space}
\end{figure}

The nonlocal term $-2\log(|\xi_\mathbf{q}|/\Lambda)$ [where $\xi(\mathbf{k}) = k_x^2 - k_y^2 - \mu$] in the Wilsonian effective action is not only nonlocal (nonanalytic) but also singular in the small $\mathbf{q}$ limit. This IR divergence in the Wilsonian effective action is puzzling as only 
high-energy modes within a finite region in momentum space are expected to contribute to the Wilsonian effective action at each step of coarse graining. In fact, this IR singularity is an artifact of using the dispersion relation $\epsilon(\mathbf{k}) = k_x^2 -k_y^2$ not only near $\mathbf{k}=0$ but also for arbitrarily large momenta. The energy dispersion $\epsilon(\mathbf{k})=k_x^2-k_y^2$ describes an infinitely extended Fermi surface, which exhibits a divergent DOS even away from the critical point due to the abundance of gapless modes on the noncompact (unbounded) Fermi surface. More specifically, as illustrated  in Fig.~[\ref{PP_Phase_Space}], the one-loop effective action exhibits the IR divergence in the particle-particle channel due to enlarging phase space for the intermediate states as the external momentum vanishes. However, real Fermi surfaces are compact, and, therefore, their finite size should be incorporated in order to avoid such a spurious singularity.~\cite{Mandal2015} To implement this, we include a momentum cutoff $K$ that suppresses contributions from modes with momenta greater than $K$ [see Eqs.~(\ref{Gamma_PP_Zmu}),(\ref{dLogLambda_Gamma_PP_Zmu})]. Although the momentum cutoff removes the singularity, we find that nonsingular yet nonlocal terms persist in the effective action unless the chemical potential is tuned away from the critical point as well. Thus, the full locality of the effective action can be kept only by introducing both a large momentum cutoff and a small chemical potential.

In general there can be two distinct sources of nonlocality in the Wilsonian effective action. nonlocality may originate from the regularization scheme, such as when a sharp (nonanalytic) cutoff in momentum space is imposed. Such nonlocalities are the artifact of the choice of the regularization scheme and can be removed by resorting to a smooth (analytic) cutoff. In contrast, nonlocality can be intrinsic, in which case it cannot be removed by choosing a smooth regularization scheme. As will be shown later, nonlocality persists in the presence of the van Hove singular point even with a smooth regularization scheme. This suggests that nonlocality at the van Hove singular point is intrinsic.

The main results of the paper are as follows. We first show that nonlocality is intrinsic in the presence of a van Hove singularity. To show this, we regularize the theory using smooth energy and momentum cutoffs. The presence of the momentum cutoff $K$ is crucial to keep the locality of the Wilsonian effective action. Treating $K$ as a dimensionful coupling constant, we capture the $\log^2 L$ growth of the attractive four-fermion contact interaction within a local RG framework away from the critical point. Note that this is in contrast with the usual $\log L$ growth of the pairing instability of regular Fermi surfaces.  Interestingly, such an enhancement of superconductivity is known to appear when a Fermi surface is coupled with a gapless boson in the context of non-Fermi liquids.~\cite{Son1999,Metlitski2014} The fast growth of the attractive interaction is present within a finite energy window, which extends all the way to the zero energy as the critical point is approached. This suggests that the system becomes unstable toward a superconducting state before the van Hove singularity is reached. We emphasize that we reach this conclusion based on a systematic local RG scheme.

The organization of this paper is as follows. We begin with the details of the model that we consider in this paper in Sec.~\ref{model}. In Sec.~\ref{RG_scheme} we lay out our RG scheme and one-loop $\beta$ functions. Section~\ref{results} contains the results of one-loop analysis, the $\beta$ functions and their implications, which are summarized in Sec.~\ref{conclusions}.


\section{Model}
\label{model}

\begin{figure}[h] 
        \centering
        \includegraphics[width=.25\textwidth]{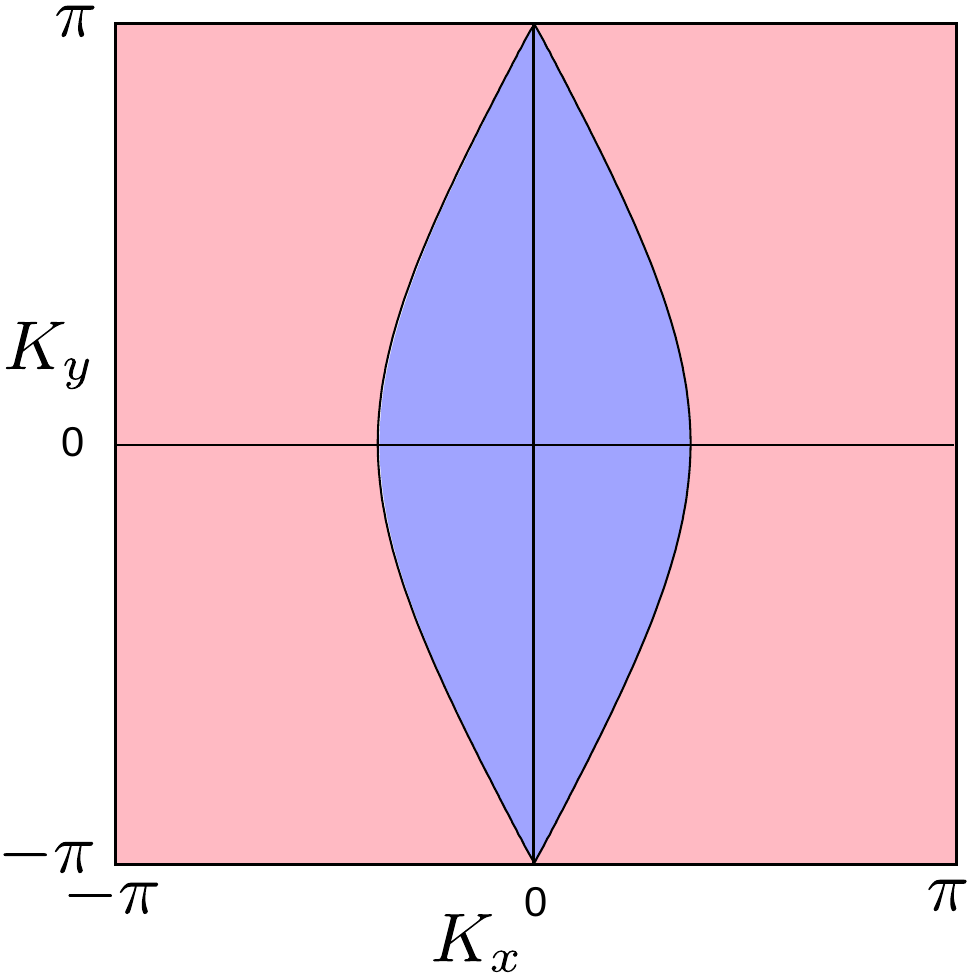}
	
        \caption[]{\justifying The Fermi surface of the lattice Hamiltonian in Eq.~(\ref{Lattice_Hamiltonian}) with the chemical potential tuned to the critical chemical potential $\mu_c=-1$. The van Hove point appears at $\mathbf{K}=(0,\pm\pi)$.}
	\label{Lattice_Model_FS}
\end{figure}

We begin by considering a lattice model that exhibits a van Hove singularity at one point in momentum space, described by the Hamiltonian,

\begin{align}
	&\mathcal{H}_{\text{\tiny Lattice}} 
	\,=\,
	\sum_{\substack{n_x,n_y\\ \sigma=\uparrow,\downarrow}}\Big[ \big\{ 
			-\frac{t_x}{2} \, c^{\dagger}_{n_x,n_y,\sigma}c_{n_x+1,n_y,\sigma}
			\, \,
	\label{Lattice_Hamiltonian}
	\\
	&
			-\frac{t_y}{2} \, c^{\dagger}_{n_x,n_y,\sigma}c_{n_x,n_y+1,\sigma}
			\,+\,
			\text{H.c.}
		\big\}
		\,-\,
		\mu \, c^{\dagger}_{n_x,n_y,\sigma}c_{n_x,n_y,\sigma}
	\Big]\,,
	\nonumber 
\end{align}
where  $t_x$ ($t_y$) is hopping amplitude in the $x$ ($y$) direction on the square lattice in two dimensions and $\mu$ is the chemical potential.
Experimentally, such a system can be realized by applying uniaxial pressure on an isotropic system, which modifies the hopping matrix elements of the corresponding tight-binding model. The above lattice Hamiltonian entails the dispersion relation $\epsilon(\mathbf{K}) \,=\, -2\cos K_x - \cos K_y$
for $t_x = 2$, $t_y = 1$ and $\mu = -1$, where $K_x$ and $K_y$ are the components of momentum measured from the center of the Brillouin zone. The resulting Fermi surface, which is shown in Fig.~[\ref{Lattice_Model_FS}], has an isolated van Hove singularity at $(0,\pm\pi)$. Expanding the dispersion relation near the singular point and rescaling the $y$-component of the momentum vector measured from the point $(0,\pm\pi)$, $\mathbf{k}$, as $k_y \rightarrow \sqrt{2} k_y$, we obtain the following quadratic saddle-point dispersion near the van Hove singular point:
\begin{equation}
	\epsilon(\mathbf{k}) \,=\, k_x^2 \,-\, k_y^2 \,+\, (\mu-\mu_c) \,+\, \mathcal{O}(k_x^4,k_y^4)\,.
\end{equation}
In fact, common van Hove singularities in 2D are all described by the quadratic saddle-point dispersion relation $\epsilon(\mathbf{k})=k_x^2-k_y^2$ near the singular point.

\begin{figure}
        \centering
        \begin{subfigure}[b]{0.22\textwidth}
                \includegraphics[scale=0.17]{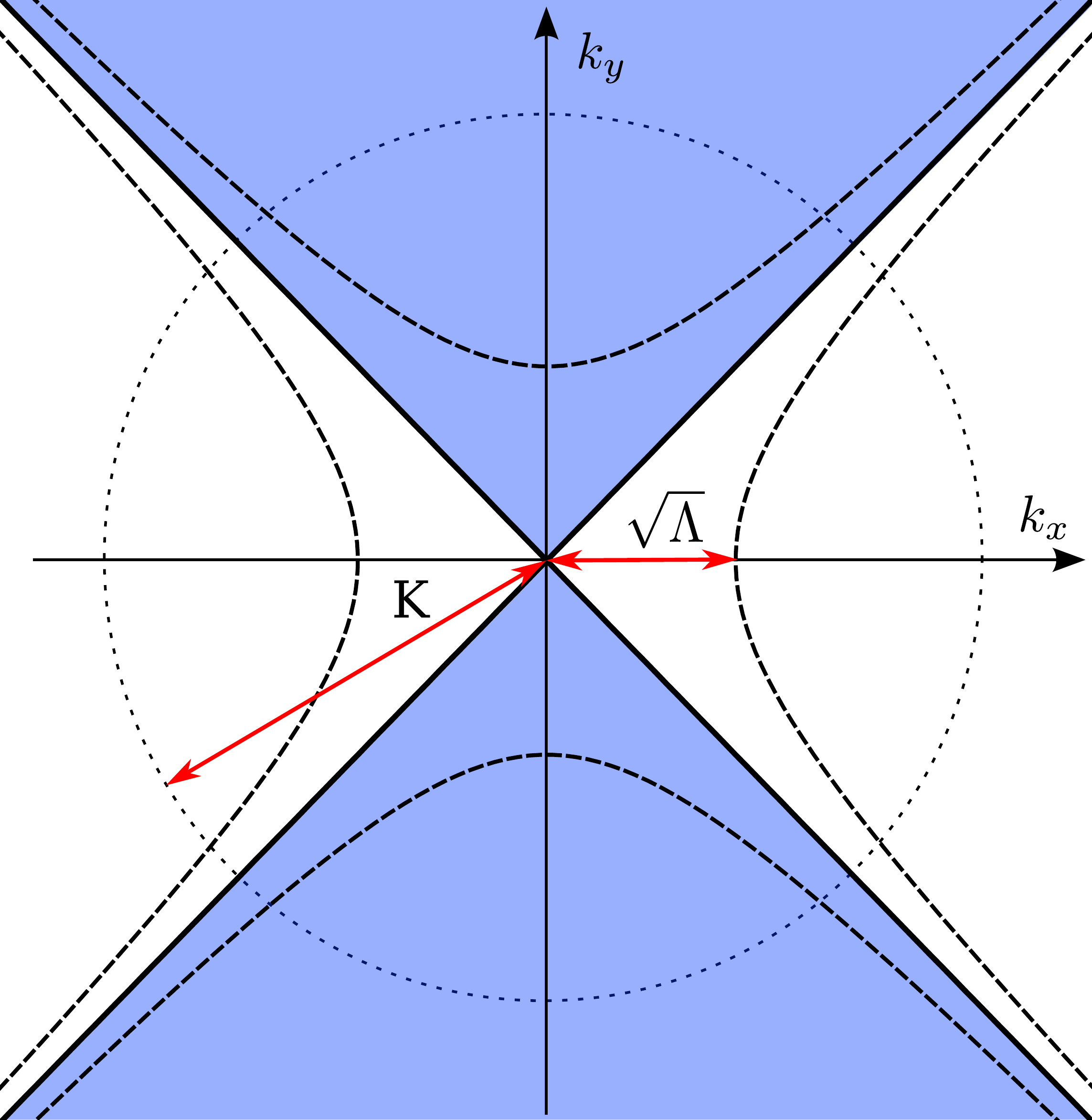}
                \caption{$\mu=0$}
                \label{VHove_FS_with_Cutoffs:Zmu}
        \end{subfigure}
	\hspace{.5cm}
        \begin{subfigure}[b]{0.22\textwidth}
                \includegraphics[scale=0.17]{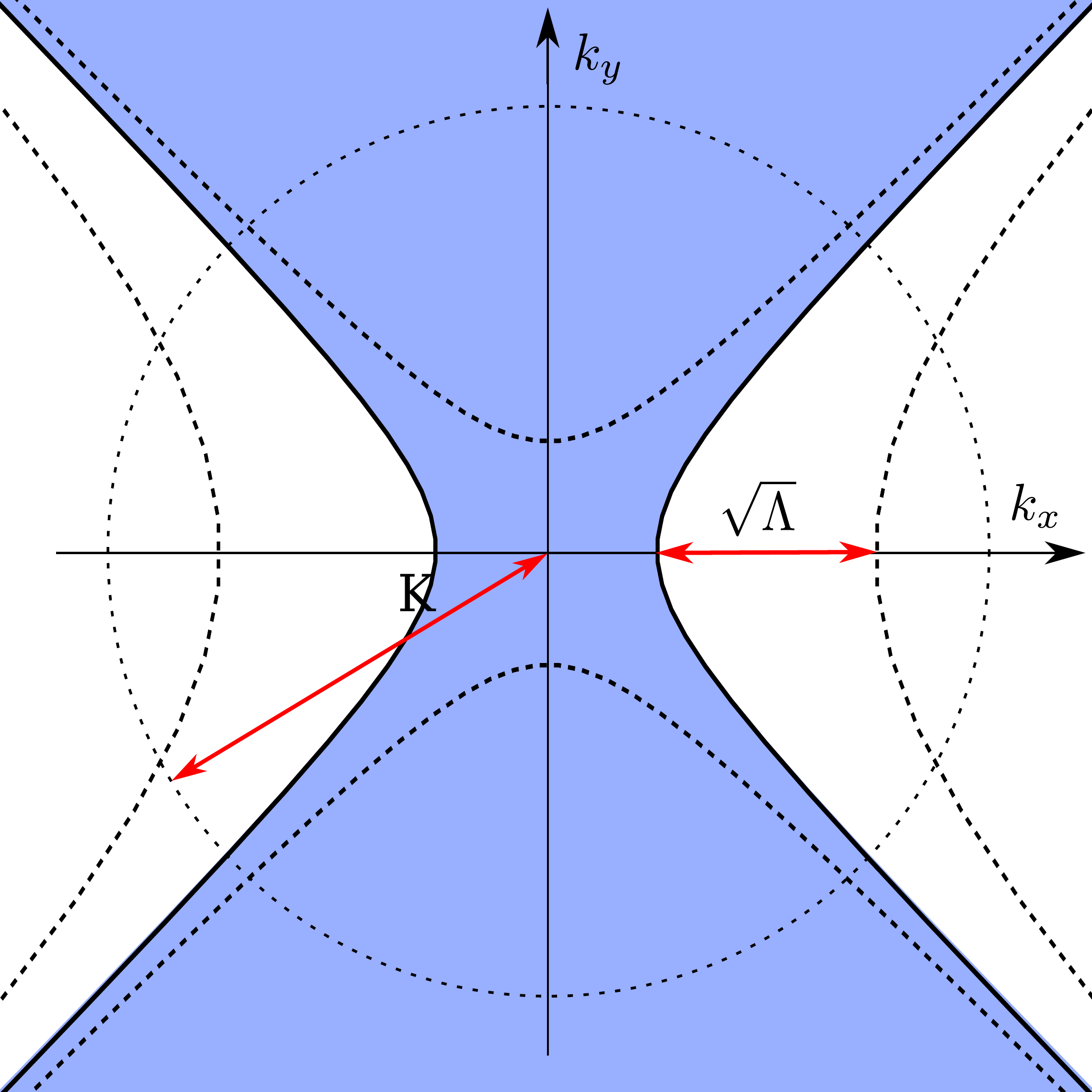}
                \caption{$\mu>0$}
                \label{VHove_FS_with_Cutoffs:NZmu}
        \end{subfigure}

        \caption[]{\justifying Depiction of the Fermi seas together with energy and momentum cutoffs at (a) the critical point of the neck-narrowing Lifshitz transition ($\mu=0$), and (b) away from the transition point ($0<\mu< \Lambda$). 
In (b) the width of the neck is $2\sqrt{\mu}$.}
	\label{VHove_FS_with_Cutoffs}
\end{figure}

The dispersion $\epsilon(\mathbf{k})=k_x^2-k_y^2$, however, suffers from a major deficiency: it does not describe a compact Fermi surface. This can be rectified either by retaining higher-order terms as in the dispersion relation $\epsilon(\mathbf{k})=k_x^2-k_y^2+\frac{1}{K^2}k_y^4$ [see Fig~\ref{Cartoon_Lifshitz_neck-narrowing}] or by imposing an explicit momentum cutoff $K$ while maintaining the quadratic dispersion near the van Hove singular point [see Fig.~\ref{VHove_FS_with_Cutoffs}]. We choose the latter scheme for the reason that, computationally, dispersions with quartic and higher-order terms are more cumbersome to deal with. In the remainder of this paper we exclusively focus on the dispersion relation $\epsilon(\mathbf{k})=k_x^2-k_y^2$ together with a momentum cutoff $K$.

Our starting point is the following regularized action:
\begin{align}
	&\mathcal{S}\,=\,\mathcal{S}_{0} + \mathcal{S}_{\text{\tiny int}}
	\label{action_model}
	\\
	&\mathcal{S}_{0}\,=\,\int\frac{\text{d}\mathbf{k}}{(2\pi)^2}\int\frac{\text{d}\omega}{2\pi} 
		\bar{\psi}_{\sigma}(\mathbf{k},\omega) \,\mathcal{G}_{0}^{-1}(\mathbf{k},\omega)\,\psi_{\sigma}(\mathbf{k},\omega)
	\nonumber \\
	&\mathcal{S}_{int}\,=\, g\Big[\prod_{i=1}^{2}\int\frac{\text{d}\mathbf{k}_i}{(2\pi)^2}\int\frac{\text{d}\omega_i}{2\pi}\Big] 
	\nonumber \\
	&\qquad \qquad 
		\bar{\psi}_{\sigma}(\mathbf{k}_1,\omega_1)
		\bar{\psi}_{\sigma'}(\mathbf{k}_2,\omega_2)
		\psi_{\sigma'}(\mathbf{k}_3,\omega_3)
		\psi_{\sigma}(\mathbf{k}_4,\omega_4)
	\nonumber \\
	&\qquad \qquad 
		\delta(\mathbf{k}_1+\mathbf{k}_2-\mathbf{k}_3-\mathbf{k}_4)\,\delta(\omega_1+\omega_2-\omega_3-\omega_4)
	\,,
	\nonumber 
\end{align}
where the partition function is given by $\mathcal{Z}=\int \mathcal{D}\psi \mathcal{D}\bar{\psi}\,e^{ -\mathcal{S} }$ and $|g|\ll1$ is the coupling of the four-fermion contact density-density interaction. Here,
\begin{equation}
	\mathcal{G}_{0}(\mathbf{k},\omega)
	=
	-\frac{e^{-\frac{\xi(\mathbf{k})^2}{\Lambda^2}}\,e^{-\frac{|\mathbf{k}|^2}{K^2}}}
	     {i\omega-\xi(\mathbf{k})}\,
	\label{G0_SoftLambdaK}
\end{equation}
is the regularized propagator, which suppresses the contributions of modes with momenta greater than $K$ \cite{Mandal2015} or energies larger than $\Lambda$, and $\xi(\mathbf{k})=k_x^2-k_y^2-\mu$. We choose to impose smooth cutoffs to maintain locality in the regularized theory.

In general, the vertex of a quartic short-range interaction term, $\Gamma_{\sigma,\sigma'}(\mathbf{k}_1,\mathbf{k}_2,\mathbf{k}_3,\mathbf{k}_4)$ ($\sigma,\sigma'=\uparrow,\downarrow$), is an analytic function. A marked difference between the RG scheme that we employ here [see Sec.~\ref{RG_scheme}] and the more conventional RG approach for regular Fermi surfaces (often referred to as Shankar's RG~\cite{Shankar1994,Polchinski1992}) is in the way that the momenta are rescaled. In Shankar's RG, only the distance from the Fermi surface is rescaled. 
Thus, for quartic short-range interactions, it is $\Gamma_{\sigma,\sigma'}(\mathbf{K}_1,\mathbf{K}_2,\mathbf{K}_3,\mathbf{K}_4)$ ($\mathbf{K}_{i}=\mathbf{k}_F$) that is taken as the marginal interaction vertex. In contrast, in our RG scheme, the momentum $\mathbf{k}$, which is measured from the van Hove point, is rescaled. Therefore, it is the leading term in the Taylor expansion of $\Gamma_{\sigma,\sigma'}(\mathbf{k}_1,\mathbf{k}_2,\mathbf{k}_3,\mathbf{k}_4)$ that gives the marginal quartic interaction vertex in our RG scheme, i.e., $g\equiv\Gamma_{\sigma,\sigma'}(0,0,0,0)$. Note that, because of the anticommutativity of fermions the spin indices should be dissimilar ($\sigma\neq\sigma'$), as $\Gamma_{\sigma,\sigma'}(0,0,0,0)$ for $\sigma=\sigma'$ reduces to a chemical potential term.

At zero chemical potential, the model described by $\epsilon(\mathbf{k})=k_x^2-k_y^2$ possesses a ``pseudo-particle-hole'' symmetry: invariance under particle-hole transformation together with a $\frac{\pi}{2}$ rotation. In the limit $K\to\infty$, this model manifests $O(1,1)$ symmetry that rotates $k_x$ into $k_y$ and vice versa with signature $(1,-1)$~\cite{Kapustin2012}. Despite the fact that the momentum cutoff alone is enough to suppress high-energy and large-momentum modes, one cannot omit the energy cutoff to treat the momentum cutoff as an energy cutoff and lower it in the course of RG. This is because lowering the momentum cutoff requires integrating out zero-energy modes (portions of the Fermi surface), which inevitably generates nonlocal terms in the Wilsonian effective action. Therefore, we treat $K$ as a dimensionful ``coupling constant'' of the theory. The RG flow is then generated by lowering $\Lambda$, which amounts to integrating out high-energy modes away from the Fermi surface.


\section{RG Scheme}
\label{RG_scheme}

\begin{figure}
        \centering
        \begin{subfigure}[b]{0.22\textwidth}
                \includegraphics[scale=0.17]{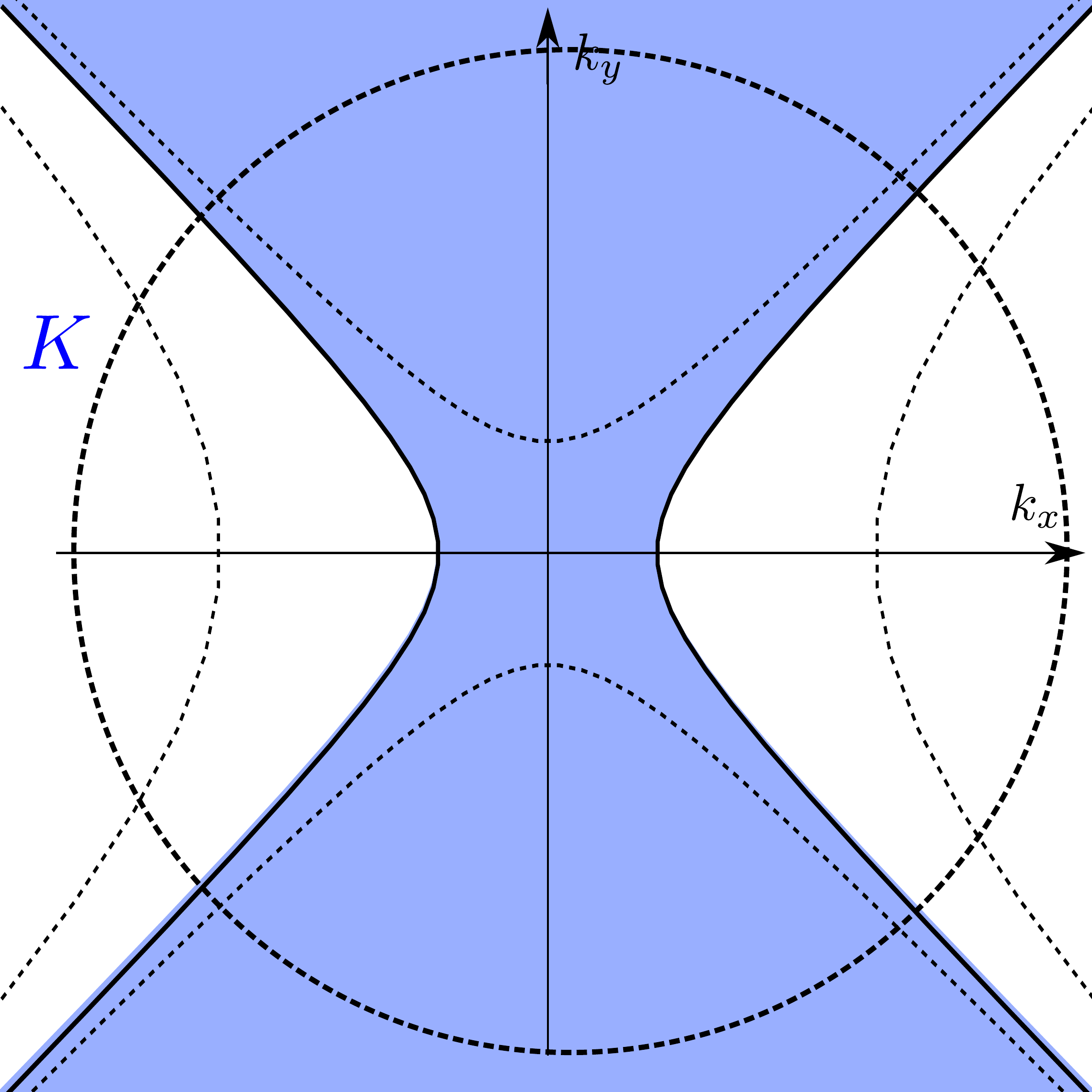}
                \caption{$\mu < \Lambda$}
                \label{NZ_VHove_FS:LLambda}
        \end{subfigure}
	\hspace{.5cm}
        \begin{subfigure}[b]{0.22\textwidth}
                \includegraphics[scale=0.17]{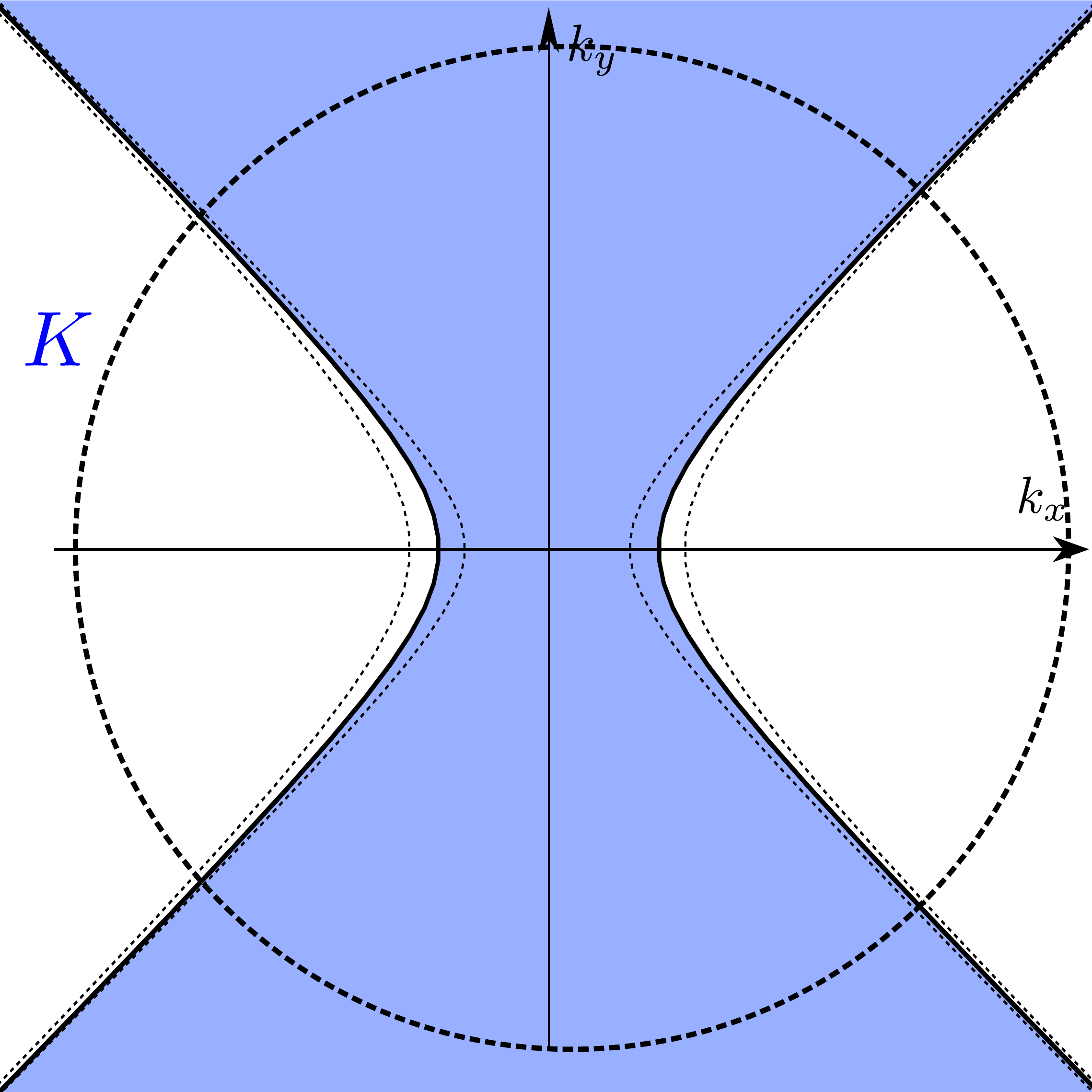}
                \caption{$\mu > \Lambda$}
                \label{NZ_VHove_FS:SLambda}
        \end{subfigure}

        \caption[]{\justifying Two regimes in lowering the energy cutoff away from the critical point of the neck-narrowing transition: (a) when $\mu < \Lambda\simeq K^2$, and (b) $\Lambda < \mu $.}
	\label{NZ_VHove_FS}
\end{figure}

We outline the renormalization group scheme in this section. Starting from the regularized action in Eq.~(\ref{action_model}), we lower the energy cutoff $\Lambda$ to $\Lambda^{\prime} = \Lambda e^{-\text{d}\ell}$. Then we add counter terms to the action so that the theory with the lowered energy cutoff reproduces the quantum effective action of the original theory. For this purpose, we first compute the quantum effective action, $\Gamma(g,\mu,K;\Lambda)$, as a function of the energy cutoff $\Lambda$ order by order in $g$. The counter term that renormalizes the Wilsonian effective action to the leading order in $g$ is then given by $\dfrac{ \partial \Gamma(g,\mu,K;\Lambda) }{ \partial \log \Lambda } \text{d}\ell$. This amounts to integrating out modes with energies between $\Lambda^{\prime}$ and $\Lambda$, which generates quantum corrections to the 
Wilsonian effective action with the new energy cutoff $\Lambda^{\prime}$. Finally, we rescale energy and momentum such that
the original energy cutoff $\Lambda$ is restored. Note that, although the quantum effective action is in general nonanalytic in external energy and momentum, it is crucial to keep the analyticity of the Wilsonian effective action.

The tree-level scalings are as follows:
\begin{equation}
	\begin{cases}
		\mathbf{k} \quad\rightarrow\quad e^{\frac{1}{2}\text{d}\ell} \,\,\, \mathbf{k}
		\\
		\mu \quad\rightarrow\quad e^{\text{d}\ell} \,\,\, \mu
		\\
		\omega \quad\rightarrow\quad e^{\text{d}\ell} \,\,\, \omega
		\\
		K \quad\rightarrow\quad e^{\frac{1}{2}\text{d}\ell} \,\,\, K
		\\
		\{\psi,\bar{\psi}\} \rightarrow e^{-\frac{3}{2}\text{d}\ell} \, \{\psi,\bar{\psi}\}
	\end{cases}
	\rightarrow
	\begin{cases}
		[\mathbf{k}] \quad=\quad \frac{1}{2}
		\\
		[\mu] \quad=\quad 1
		\\ 
		[\omega] \quad=\quad 1
		\\
		[K] \quad=\quad \frac{1}{2}
		\\
		[\psi]=[\bar{\psi}] = -\frac{3}{2}
	\end{cases}
\end{equation}
From the above tree-level scaling dimensions we obtain $[g] = 0$, and thus the four-fermion interaction with momentum-independent vertex is marginal.
It is noted that the momentum cutoff $K$ and the chemical potential $\mu$ run under the rescaling. 
As a result, these parameters should be treated as relevant couplings in the theory.~\cite{Mandal2015}

\begin{figure}
        \centering
  	\vspace{1cm}
        \begin{subfigure}[b]{0.15\textwidth}
		 \begin{fmfgraph*}(400,200)
			\fmfleft{i1,i2}
			\fmfright{o1,o2}  
			\fmflabel{$\mathbf{k}_{_4},\sigma$}{i1}
			\fmflabel{$\mathbf{k}_{_1},\sigma$}{i2}
			\fmflabel{$\mathbf{k}_{_3},\sigma$}{o1}
			\fmflabel{$\mathbf{k}_{_2},\sigma$}{o2}
			\fmf{fermion}{i1,v1,i2}
			\fmf{fermion}{o1,v4,o2}
			\fmf{photon,label=$g$}{v1,v2}	
			\fmf{fermion,left,tension=0.35,label=$\sigma'$}{v2,v3}
			\fmf{fermion,left,tension=0.35,label=$\sigma'$}{v3,v2}
			\fmf{photon,label=$g$}{v3,v4}
			\fmfdotn{v}{4}
		\end{fmfgraph*}
		\vspace{.25cm}
                \caption{$\Pi_{\text{\tiny PH}}$}
                \label{1loop_interaction:PH1}
        \end{subfigure}
	\hspace{1.5cm}
        \begin{subfigure}[b]{0.15\textwidth}
		\begin{fmfgraph*}(400,200)
			\fmfleft{i1,i2}
			\fmfright{o1,o2}  
			\fmflabel{$\mathbf{k}_{_4}$}{i1}
			\fmflabel{$\mathbf{k}_{_1}$}{i2}
			\fmflabel{$\mathbf{k}_{_3}$}{o1}
			\fmflabel{$\mathbf{k}_{_2}$}{o2}
			\fmf{fermion}{i1,v1}
			\fmf{fermion,label=$$,label.side=right}{v1,v3}
			\fmf{fermion,label=$$,label.side=right}{v3,v2}
			\fmf{fermion}{v2,i2}
			\fmf{fermion}{o1,v4,o2}
			\fmf{photon,tension=0,label=$g$,label.side=left}{v1,v2}
			\fmf{photon,label=$g$}{v3,v4}
			\fmfdotn{v}{4}
  		\end{fmfgraph*}
		\vspace{.25cm}
                \caption{$\Gamma^{\text{\tiny Penguin}}_{\text{\tiny PH}}$}
                \label{1loop_interaction:PH2}
        \end{subfigure}
	
	\vspace{1.cm}
        \begin{subfigure}[b]{0.15\textwidth}                
  		\begin{fmfgraph*}(400,200)
			\fmfleft{i1,i2}
			\fmfright{o1,o2}  
			\fmflabel{$\mathbf{k}_{_4},\sigma$}{i1}
			\fmflabel{$\mathbf{k}_{_2},\sigma'$}{i2}
			\fmflabel{$\mathbf{k}_{_1},\sigma$}{o1}
			\fmflabel{$\mathbf{k}_{_3},\sigma'$}{o2}
			\fmf{fermion}{i1,v1}
			\fmf{fermion,label=$$}{v1,v4}
			\fmf{fermion}{v4,o1}
			\fmf{fermion}{o2,v3}
			\fmf{fermion,label=$$}{v3,v2}
			\fmf{fermion}{v2,i2}
			\fmf{photon,tension=0,label=$g$,label.side=left}{v1,v2}
			\fmf{photon,tension=0,label=$g$,label.side=left}{v3,v4}
			\fmfdotn{v}{4}
  		\end{fmfgraph*}
		\vspace{.25cm}
                \caption{$\Gamma^{\text{\tiny Ladder}}_{\text{\tiny PH}}$}
                \label{1loop_interaction:PH3}
        \end{subfigure}
	\hspace{1.5cm}
        \begin{subfigure}[b]{0.15\textwidth}       
  		\begin{fmfgraph*}(400,200)
			\fmfleft{i1,i2}
			\fmfright{o1,o2}  
			\fmflabel{$\mathbf{k}_{_4}$}{i1}
			\fmflabel{$\mathbf{k}_{_3}$}{i2}
			\fmflabel{$\mathbf{k}_{_1}$}{o1}
			\fmflabel{$\mathbf{k}_{_2}$}{o2}
			\fmf{fermion}{i1,v1}
			\fmf{fermion,label=$$}{v1,v3}
			\fmf{fermion}{v3,o1}
			\fmf{fermion}{i2,v2}
			\fmf{fermion,label=$$,label.side=left}{v2,v4}
			\fmf{fermion}{v4,o2}
			\fmf{photon,tension=0,label=$g$,label.side=left}{v1,v2}
			\fmf{photon,tension=0,label=$g$,label.side=right}{v3,v4}
			\fmfdotn{v}{4}
  		\end{fmfgraph*}
		\vspace{.25cm}
                \caption{$\Gamma_{\text{\tiny PP}}$}
                \label{1loop_interaction:PP}
        \end{subfigure}

        \caption[]{\justifying Diagrams that renormalize four-fermion interactions at one-loop order. In order to keep track of the spin indices, a wiggly line is used for the marginal density-density interaction even though the vertex of this interaction is momentum-independent. Diagrams (a), (b) and (c) involve the exchange of a particle and a hole, whereas diagram (d) is a particle-particle diagram. Diagram (a) is the usual particle-hole bubble $\Pi_{\text{\tiny PH}}$, diagram (b) is the penguin diagram, diagram (c) is a one-loop ladder diagram $\Gamma_{\text{\tiny PH}}^{\text{\tiny Ladder}}$.}
	\label{1loop_interaction}
\end{figure}
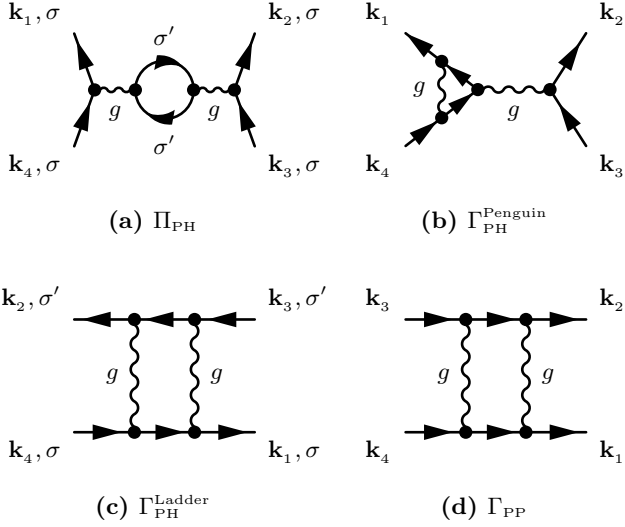

\begin{figure}
        \centering
	\vspace{1cm}
        \begin{subfigure}[b]{0.22\textwidth}              
  		\begin{fmfgraph*}(400,200)
			\fmfleft{i}
			\fmfright{o}  
			\fmftop{v2}
			\fmflabel{$\mathbf{q},\Omega$}{i}
			\fmf{fermion}{i,v1}
			\fmf{fermion}{v1,o}
			\fmf{photon,tension=0,label=$g$}{v1,v2}	
			\fmf{fermion,tension=0.7,label=$\mathbf{k}$,label.side=left}{v2,v2}
			\fmfdotn{v}{2}
  		\end{fmfgraph*}
		\vspace{.25cm}
                \caption{$\Sigma^{(1)}$}
                \label{1loop-Sigma:1}
        \end{subfigure}
	\hspace{.5cm}
        \begin{subfigure}[b]{0.22\textwidth}
  		\begin{fmfgraph*}(400,200)
			\fmfleft{i}
			\fmfright{o}  
			\fmflabel{$\mathbf{q},\Omega$}{i}
			\fmf{fermion}{i,v1}
			\fmf{fermion,label=$\mathbf{k}$}{v1,v2}
			\fmf{fermion}{v2,o}
			\fmf{photon,left,tension=0,label=$g$,label.side=left}{v1,v2}
			\fmfdotn{v}{2}
  		\end{fmfgraph*}
		\vspace{.25cm}
                \caption{$\Sigma^{(1)}$}
                \label{1loop-Sigma:2}
        \end{subfigure}

        \caption[]{\justifying Self-energy diagrams at one-loop order. Since we are considering momentum-independent bare interactions, the Fock diagram, (b), is not allowed.}
	\label{1loop-Sigma}
\end{figure}
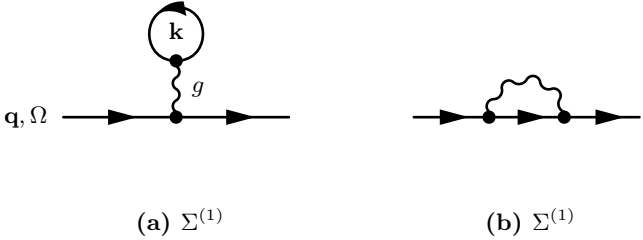

At one-loop order, we have diagrams shown in Figs.~[\ref{1loop_interaction},\ref{1loop-Sigma}]. Among the diagrams in Fig.~[\ref{1loop_interaction}] that renormalize the interaction vertex, the diagram in Fig.~[\ref{1loop_interaction:PH2}] is not allowed as it requires same spin indices on all four legs of one of the vertices. Diagrams in Fig.~[\ref{1loop_interaction:PH1}] and Fig.~[\ref{1loop_interaction:PH3}] do not vanish and involve the exchange of a particle and a hole, while the diagram in Fig.~[\ref{1loop_interaction:PP}] involves a pair of particles. Note that the results of the diagrams in Fig.~[\ref{1loop_interaction:PH1}] and Fig.~[\ref{1loop_interaction:PH3}] to the interaction vertex are distinct due to the spin indices of the external legs; nevertheless they have the same loop integral up to a minus sign:
\begin{align}
	&
	\Gamma_{\text{\tiny PH}}^{\text{\tiny Ladder}}(\mathbf{q},\Omega)
	\,=\,
	-\Pi_{\text{\tiny PH}}(\mathbf{q},\Omega)
	\label{1-loop_PH} 
	\\[.5em]
	&\,=\,
	\int\frac{\text{d}\mathbf{k}}{(2\pi)^2}\,
	\frac{\theta(\xi_{\mathbf{k}}) - \theta(\xi_{\mathbf{k+q}})}
	     {i\Omega - \xi(\mathbf{k+q})+\xi(\mathbf{k})}
	e^{-\frac{\xi_{\mathbf{k}}^2+\xi_{\mathbf{k+q}}^2}{\Lambda^2}}
	e^{-\frac{|\mathbf{k}|^2+|\mathbf{k+q}|^2}{K^2}}
	\,.
	\nonumber
\end{align}
The result of the diagram in Fig.~[\ref{1loop_interaction:PP}] is given by:
\begin{align}
	\Gamma_{\text{\tiny PP}}(\mathbf{q},\Omega)
	\,=\,&
	-\int\frac{\text{d}\mathbf{k}}{(2\pi)^2}\,
	\frac{\theta(\xi_{\mathbf{k}}) - \theta(-\xi_{\mathbf{k+q}})}
	     {i\Omega - \xi(\mathbf{k+q})-\xi(\mathbf{k})}\,
	\nonumber \\
	&
	e^{-\frac{\xi_{\mathbf{k}}^2+\xi_{\mathbf{k+q}}^2}{\Lambda^2}}
	e^{-\frac{|\mathbf{k}|^2+|\mathbf{k+q}|^2}{K^2}}\,.
	\label{1-loop_PP}
\end{align}
The renormalized interaction term at one-loop order is given by,
\begin{align}	
	\parbox{15mm}{
 		\begin{fmfgraph*}(200,100)
			\fmfset{arrow_len}{2mm} 
			\fmfset{arrow_ang}{20} 
			\fmfset{dot_size}{1mm}
			\fmfset{wiggly_len}{2mm}
			\fmfleft{i1,i2}
			\fmfright{o1,o2} 
			\fmfrpolyn{shade,label={\tiny $\Gamma^{(4)}$},label.dist=3mm}{G}{4}
			\fmf{fermion}{i1,G1}
			\fmf{fermion}{G2,i2}
			\fmf{fermion}{o1,G4}
			\fmf{fermion}{G3,o2}	
 		\end{fmfgraph*}
	}
	\,=\,
	&
	g
	\parbox{15mm}{
 		\begin{fmfgraph*}(200,100)
			\fmfset{arrow_len}{2mm} 
			\fmfset{arrow_ang}{20} 
			\fmfset{dot_size}{1mm}
			\fmfset{wiggly_len}{2mm}
			\fmfleft{i1,i2}
			\fmfright{o1,o2}  
			\fmf{fermion}{i1,v1,i2}
			\fmf{fermion}{o1,v2,o2}
			\fmf{photon}{v1,v2}
			\fmfdotn{v}{2}
 		\end{fmfgraph*}
	}
	\,\,-\,\,
	\frac{g^2}{2}
	\parbox{15mm}{
 		\begin{fmfgraph*}(200,100)
			\fmfset{arrow_len}{2mm} 
			\fmfset{arrow_ang}{20}
			\fmfset{dot_size}{1mm}
			\fmfset{wiggly_len}{2mm}
			\fmfleft{i1,i2}
			\fmfright{o1,o2}  
			\fmf{fermion}{i1,v1,i2}
			\fmf{fermion}{o1,v4,o2}
			\fmf{photon}{v1,v2}	
			\fmf{fermion,left,tension=0.35}{v2,v3}
			\fmf{fermion,left,tension=0.35}{v3,v2}
			\fmf{photon}{v3,v4}
			\fmfdotn{v}{4}
 		\end{fmfgraph*}
	}
	\\[1em]
	&	
	-
	g^2
	\parbox{15mm}{
		\begin{fmfgraph*}(200,100)
			\fmfset{arrow_len}{2mm} 
			\fmfset{arrow_ang}{20}
			\fmfset{wiggly_len}{2mm}
			\fmfset{dot_size}{1mm}
			\fmfleft{i1,i2}
			\fmfright{o1,o2}  
			\fmf{fermion}{i1,v1}
			\fmf{fermion}{v1,v4}
			\fmf{fermion}{v4,o1}
			\fmf{fermion}{o2,v3}
			\fmf{fermion}{v3,v2}
			\fmf{fermion}{v2,i2}
			\fmf{photon,tension=0}{v1,v2}
			\fmf{photon,tension=0}{v3,v4}
			\fmfdotn{v}{4}
		\end{fmfgraph*}
	}
	\,\,-\,\,
	g^2
	\parbox{15mm}{
		\begin{fmfgraph*}(200,100)
			\fmfset{arrow_len}{2mm} 
			\fmfset{arrow_ang}{20}
			\fmfset{dot_size}{1mm}
			\fmfset{wiggly_len}{2mm}
			\fmfleft{i1,i2}
			\fmfright{o1,o2}  
			\fmf{fermion}{i1,v1}
			\fmf{fermion}{v1,v3}
			\fmf{fermion}{v3,o1}
			\fmf{fermion}{i2,v2}
			\fmf{fermion,label.side=left}{v2,v4}
			\fmf{fermion}{v4,o2}
			\fmf{photon,tension=0}{v1,v2}
			\fmf{photon,tension=0}{v3,v4}
			\fmfdotn{v}{4}
  		\end{fmfgraph*}
	}
	\nonumber
	\,,
\end{align}
where the accompanying numerical factors $-\frac{1}{2}$, $-1$ and $-1$ are the combinatorial factors associated with each diagram. We postpone the discussion of the structure of $\partial_{\log\Lambda}\Gamma^{\text{\tiny Ladder}}_{\text{\tiny PH}}(\mathbf{q},\Omega=0)$ and $\partial_{\log\Lambda}\Gamma_{\text{\tiny PP}}(\mathbf{q},\Omega=0)$ to the next section. Note that we are solely interested in the quantum corrections at zero frequency.

Diagrams shown in Fig.~[\ref{1loop-Sigma}] contribute to the renormalization of the chemical potential at one-loop order. The pseudo-particle-hole symmetry at $\mu=0$ ensures that the contribution of these diagrams (as well as all higher-order corrections to the chemical potential) vanishes at $\mu=0$. For $\mu>0$, $\partial_{\log\Lambda}\Sigma^{(1)} \propto g\mu$ (with no singular dependence on $K$ as $K\to\infty$). For this reason, we ignore the quantum correction to the chemical potential in the small $g$ and $\mu$ limit near the critical point. Since we assume that the bare interaction vertex is momentum independent, the momentum-dependent self-energy contribution first appears at the two-loop order.

When quantum corrections are analytic, they can be expanded in the basis of local operators. Taylor expanding the vertices of one-loop four-fermion quantum corrections, one obtains the following local interaction terms:
\begin{align}
	\mathcal{S}_{\text{\tiny PH}}^{\{n,m\}} =
	& 
	\Big[
		g^{\{n,m\}}_{\text{\tiny PH}}(1-\delta_{\sigma,\sigma'}) + h^{\{n,m\}}_{\text{\tiny PH}}\delta_{\sigma,\sigma'}
	\Big]
	\nonumber \\
	&
	\int\frac{\text{d}\mathbf{k}}{(2\pi)^2} \frac{\text{d}\omega}{2\pi} 
	\int\frac{\text{d}\mathbf{k}'}{(2\pi)^2}\frac{\text{d}\omega'}{2\pi}
	\int\frac{\text{d}\mathbf{q}}{(2\pi)^2}\frac{\text{d}\Omega}{2\pi}
	\nonumber \\
	&
	\bar{\psi}_{\sigma,\mathbf{k}}\,\psi_{\sigma,\mathbf{k+q}}\,
	\,\frac{q_x^{2n}}{\Lambda^{n}}\,\frac{q_y^{2m}}{\Lambda^{m}}\,
	\bar{\psi}_{\sigma',\mathbf{k'+q}}\,\psi_{\sigma',\mathbf{k'}}\,,
	\label{Irrelevant_PH_Interaction}	
\end{align}
which are the local particle-hole interactions with the couplings $g^{\{n,m\}}_{\text{\tiny PH}}$ ($h^{\{n,m\}}_{\text{\tiny PH}}$ for similar spin indices). Similarly,
\begin{align}
	\mathcal{S}_{\text{\tiny PP}}^{\{n,m\}} =
	&
	g^{\{n,m\}}_{\text{\tiny PP}}
	\int\frac{\text{d}\mathbf{k}}{(2\pi)^2} \frac{\text{d}\omega}{2\pi} 
	\int\frac{\text{d}\mathbf{k}'}{(2\pi)^2}\frac{\text{d}\omega'}{2\pi}
	\int\frac{\text{d}\mathbf{q}}{(2\pi)^2}\frac{\text{d}\Omega}{2\pi}
	\nonumber \\
	&
	\bar{\psi}_{\sigma',\mathbf{k+q}}\,\bar{\psi}_{\sigma,\mathbf{-k}}\,
	\,\frac{q_x^{2n}}{\Lambda^{n}}\,\frac{q_y^{2m}}{\Lambda^{m}}\,
	\psi_{\sigma,\mathbf{k'+q}}\,\psi_{\sigma',\mathbf{-k'}}\,
	\label{Irrelevant_PP_Interaction}	
\end{align}
are the particle-particle interactions with the couplings $g^{\{n,m\}}_{\text{\tiny PP}}$. Here $n$ and $m$ are respectively the powers of $q_x^2$ and $q_y^2$ in the Taylor expansions of the one-loop quantum corrections. The one-loop $\beta$ functions for the above local operators are as follows:
\begin{align}
	&\dot{g}\,=\, - g^2\,\partial_{\log\Lambda} \Gamma^{\text{\tiny Ladder}}_{\text{\tiny PH}}(0) 
		      - g^2\,\partial_{\log\Lambda} \Gamma_{\text{\tiny PP}}(0)
	\nonumber \\
	&\dot{g}^{\{n,m\}}_{\text{\tiny PH}} = -(n+m)\,g^{\{n,m\}}_{\text{\tiny PH}} 
				- g^2\frac{\partial^{2n}}{\partial q_x^{2n}}\frac{\partial^{2m}}{\partial q_y^{2m}}
				  \partial_{\log\Lambda} \Gamma^{\text{\tiny Ladder}}_{\text{\tiny PH}}(\mathbf{q})\big|_{0} 
	\nonumber \\
	&\dot{h}^{\{n,m\}}_{\text{\tiny PH}} = -(n+m)\,h^{\{n,m\}}_{\text{\tiny PH}} 
				- \frac{g^2}{2} \,\frac{\partial^{2n}}{\partial q_x^{2n}}\frac{\partial^{2m}}{\partial q_y^{2m}}
				  \partial_{\log\Lambda} \Pi_{\text{\tiny PH}}(\mathbf{q}) \big|_{0} 
	\nonumber \\
	&\dot{g}^{\{n,m\}}_{\text{\tiny PP}} = -(n+m)\,g^{\{n,m\}}_{\text{\tiny PP}} 
				- g^2\frac{\partial^{2n}}{\partial q_x^{2n}}\frac{\partial^{2m}}{\partial q_y^{2m}}
				  \partial_{\log\Lambda}\Gamma_{\text{\tiny PP}}(\mathbf{q})\big|_{0}
	\nonumber \\
	&\dot{K}\,=\, \frac{1}{2}K
	\nonumber \\
	&\dot{\mu}\,=\, \mu + \mathcal{O}(g)\,,
\label{Formal-beta_functions}
\end{align}
where the vertical bar with the subscript $0$ is a shorthand for $|_{q=0}$, and the numerical factors accompanying $g^2$ terms are the combinatorial factors associated with the corresponding diagrams. Here, $g$ is the coupling of the momentum-independent interaction term.

\section{One-Loop RG Analysis}
\label{results}

We explicitly compute one-loop $\beta$ functions in this section. We specifically demonstrate that both a nonzero chemical potential and a finite momentum cutoff are needed to maintain the locality of the Wilsonian effective action. When the Wilsonian effective action is Taylor expanded in momentum, the radius of convergence shrinks to zero either when $\mu$ diminishes or $K$ grows large. To simplify the computation, we first focus on one-loop particle-hole quantum corrections ($\partial_{\log\Lambda}\Gamma_{\text{\tiny PH}}$ -- from here on we use $\Gamma_{\text{\tiny PH}}\equiv\Gamma_{\text{\tiny PH}}^{\text{\tiny Ladder}}=-\Pi_{\text{\tiny PH}}$) for $\mathbf{q}=q\hat{x}$.


\subsection{$0<\mu\ll \Lambda$} 

\begin{figure}
        \centering
        \includegraphics[scale=0.2]{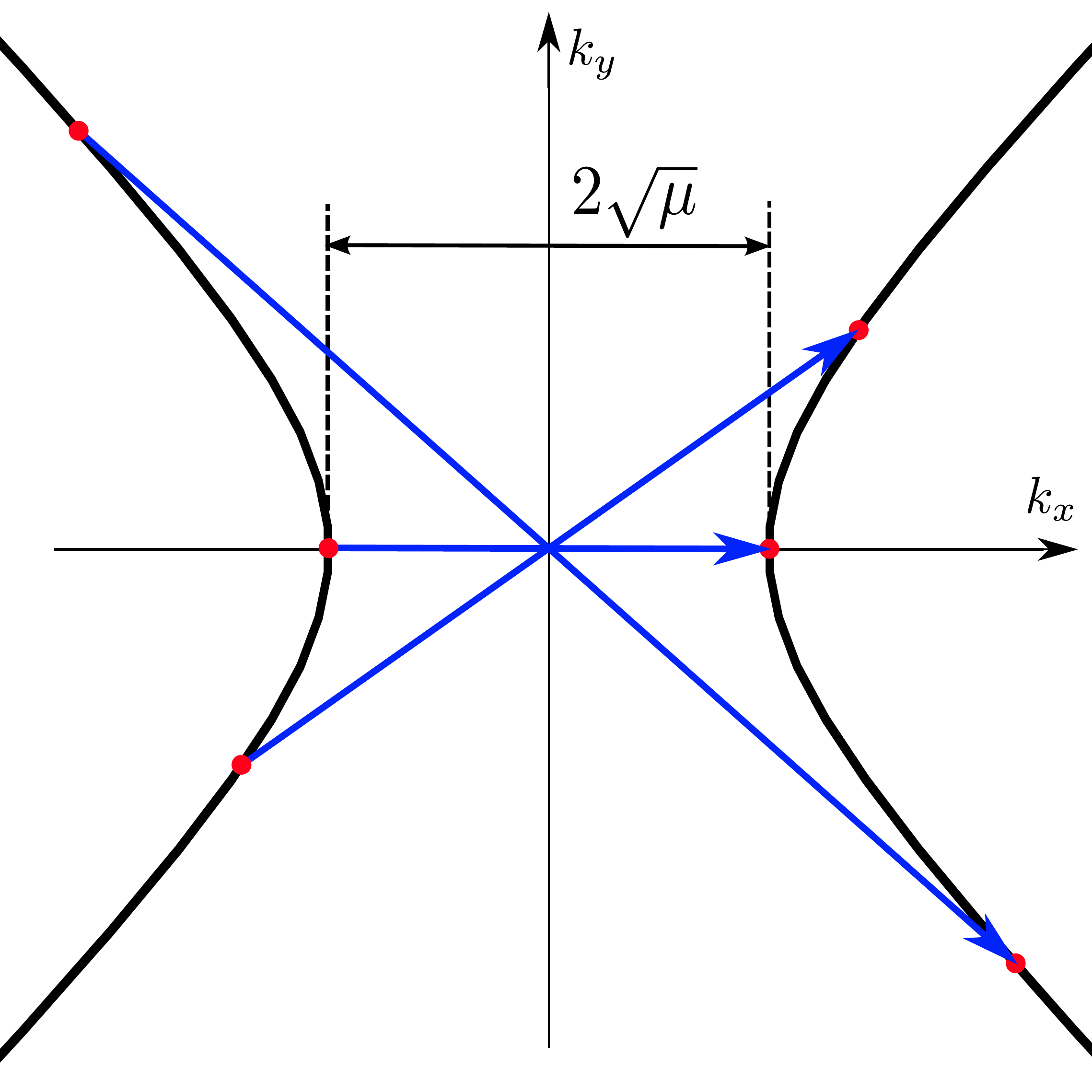}
        \caption[]{\justifying All three indicated vectors satisfying $q_x^2-q_y^2=4\mu$ and connecting antipodal points correspond to ``$2k_F$'' momentum transfers $\mathbf{q}$ on the Fermi surface.}
	\label{2kF}
\end{figure}

When the chemical potential is nonzero,
the one-loop correction from the particle-hole diagram
is given by:
\begin{align}
	&\partial_{\log \Lambda}\Gamma_{\text{\tiny PH}} (q\hat{x},0)
	\approx
	\frac{1}{2(2\pi)^2}
	\Bigg[
		8\frac{K^2}{\Lambda}\frac{q^2}{\Lambda}
		+
		\Big(
			19.7\log\frac{\mu}{K^2}  
	\label{PH_NZmu_qx_l2kF}	
	\\
	&  
			+
			101
			-
			5.6\frac{K^4}{\Lambda^2} 
		\Big)\frac{q^4}{\Lambda^2}
		+
		\Big(
			-
			1.87\frac{\Lambda}{\mu}
			+
			\big[
				22.8\frac{K^2}{\Lambda}
				+
				86\frac{\Lambda}{K^2}
			\big]\times
	\nonumber \\
	& 
			\log\frac{\mu}{K^2}
			+
			4.11\frac{K^6}{\Lambda^3} 
			+
			162\frac{K^2}{\Lambda}
		\Big)\frac{q^6}{\Lambda^3}
		+
		\Big(
			-
			0.13\frac{\Lambda^2}{\mu^2}
	\nonumber \\
	& 
			-
			2.06\frac{K^2}{\mu}
			+
			\big[
				237\frac{\Lambda^2}{K^4}
				+
				255
				+
				8.7\frac{K^4}{\Lambda^2}
			\big]\log\frac{\mu}{K^2}
			+
			1462
	\nonumber \\
	& 
			-
			71.5\frac{K^4}{\Lambda^2}
			-
			3.16\frac{K^8}{\Lambda^4}
		\Big)\frac{q^8}{\Lambda^4}
		+
		\Big(
			-
			0.014\frac{\Lambda^3}{\mu^3}
			-
			0.138\frac{K^2}{\Lambda\mu^2}
	\nonumber \\
	& 
			-
			0.326\frac{\Lambda^3}{K^2\mu^2}
			-
			0.716\frac{K^4}{\Lambda\mu}
			-
			17.13\frac{\Lambda}{\mu}
			-
			12.98\frac{\Lambda^3}{K^4\mu}
	\nonumber \\
	& 
			+\big[
				-
				1.262\frac{K^6}{\Lambda^3} 
				+
				305\frac{K^2}{\Lambda}
				+
				1082\frac{\Lambda}{K^2}
				+
				483\frac{\Lambda^3}{K^6}
			\big]\log\frac{\mu}{K^2} 
	\nonumber \\
	& 
			+
			2.52\frac{K^{10}}{\Lambda^5}
			+
			9.235\frac{K^6}{\Lambda^3}
			+
			1987\frac{K^2}{\Lambda}
		\Big)\frac{q^{10}}{\Lambda^5}
		+
		\mathcal{O}(q)^{12}
	\Bigg]
	\nonumber
	\,,
\end{align}
where we have ignored all $\mathcal{O}(\mu)$ and $\mathcal{O}(1/K)$ terms.

\begin{figure}
        \centering
        \includegraphics[scale=0.2]{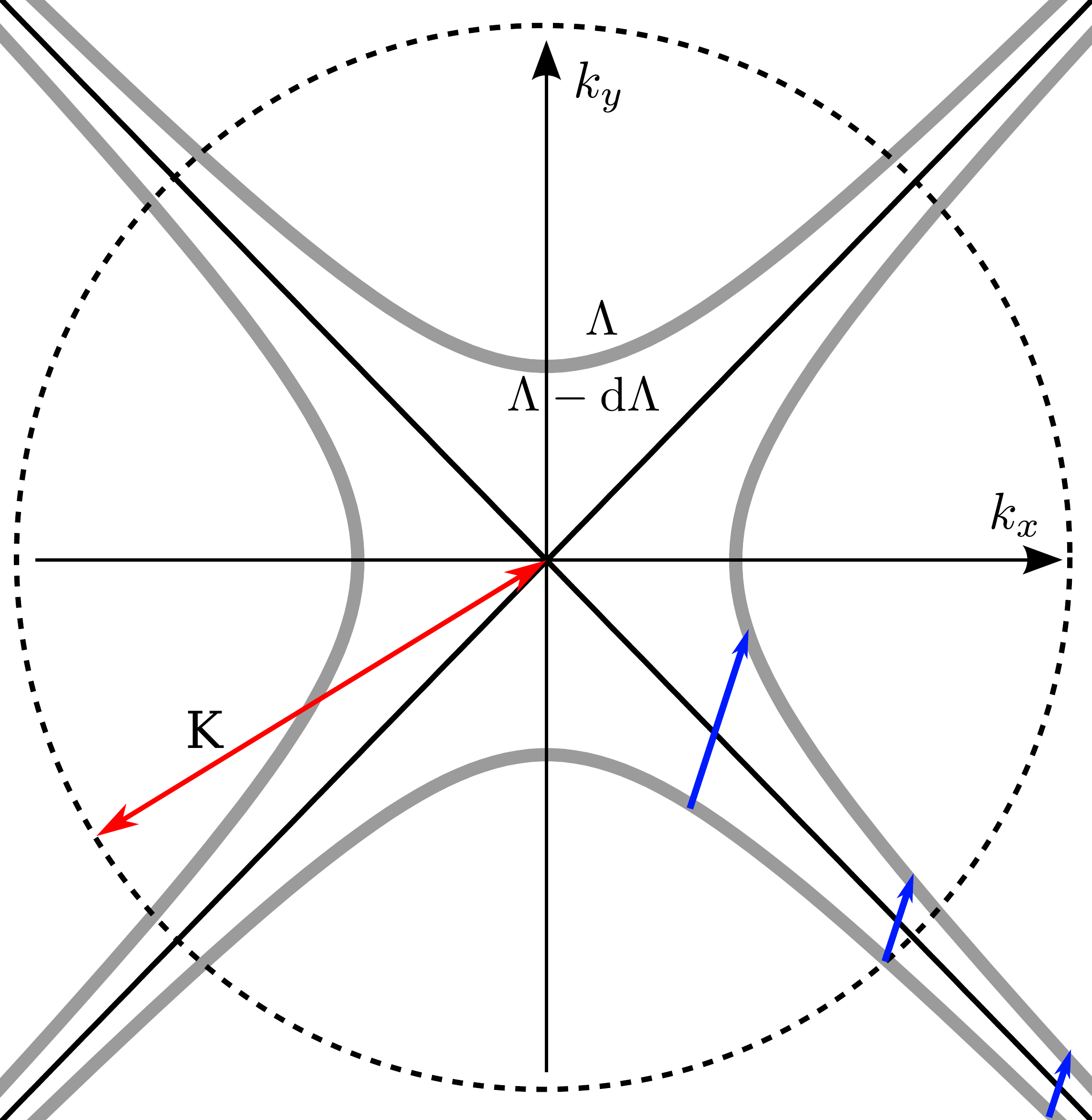}
         
        \caption[]{\justifying When both cutoffs are imposed sharply, as depicted in this figure, for smaller $|\mathbf{q}|$, the eliminated modes will be farther from the origin. 
Thus, if $|\mathbf{q}|$ is sufficiently small ($<\frac{\Lambda}{K}$) $\partial_{\log \Lambda}\Gamma_{\text{\tiny PH}} (\mathbf{q},\Omega)$ suddenly vanishes (as a result of using sharp cutoffs). When both cutoffs are imposed softly, as discussed in the text, this feature results in the convergence radius $q< \Lambda/K$ in $\partial_{\log \Lambda}\Gamma_{\text{\tiny PH}}$.}
	\label{VHove_Cutoffs}
\end{figure}

The above expression is valid for $q^2 < 4 \mu$. The nonanalyticity at $q^2=4\mu$ stems from the $2k_F$ singularity, which arises when the transfer momentum $\mathbf{q}$ connects antipodal points on the Fermi surface, as shown in Fig.~[\ref{2kF}]. Having this in mind, let us examine the origin of different $K$- and $\mu$-dependent terms in this series expansion. First consider terms that become singular in the $\mu \to 0$ limit. The appearance of these singular terms can be understood as follows. With a soft-energy cutoff, although quantum corrections are most sensitive to modes at the energy scale $\Lambda$, they nevertheless weakly sense all other modes. This is because the derivative of a soft-energy regulator with respect to $\log\Lambda$ is not a $\delta$-function. Thus, the appearance of an IR singularity upon setting $\mu$ to zero is ``sensed'' by quantum corrections (in this case, $\partial_{\log \Lambda}\Gamma_{\text{\tiny PH}}$). Since the above series expansion is valid only for $q<2\sqrt{\mu}$ 
and singular terms in $\mu$ are accompanied by sufficiently high powers of $q$, these singular terms do not result in divergence in the $\mu \rightarrow 0$ limit. The singular dependence on $\mu$ should be understood as a sign of nonanalyticity in the $\mu = 0$ limit instead of actual divergence. Singular dependence on $\mu$ first appears at $\mathcal{O}(q^4)$. This feature depends on the choice of the energy cutoff. If $\exp(-\xi_{\mathbf{k}}^4/\Lambda^4)$ was used instead of $\exp(-\xi_{\mathbf{k}}^2/\Lambda^2)$, singular dependence on $\mu$ would first appear at $\mathcal{O}(q^6)$. More generally, for an energy cutoff of the form $\exp(-\xi_{\mathbf{k}}^{2n}/\Lambda^{2n})$, singular terms in $\mu$ appear at $\mathcal{O}(q^{2n+2})$. Note that, in the limit $n\to\infty$, where the energy cutoff becomes the sharp energy cutoff $\theta(1-\frac{|\xi_{\mathbf{k}}|}{\Lambda})$, these particular singular terms disappear. However, the sharp cutoff will generate yet another nonanalyticity in $\partial_{\log\
Lambda}\Gamma_{\text{\tiny PH}}$ at $|\mathbf{q}|\simeq\Lambda/K$, which is pushed to zero as RG progresses [see Fig.~\ref{VHove_Cutoffs}].

Another important feature of the series expansion in Eq.~(\ref{PH_NZmu_qx_l2kF}) is the presence of terms with positive powers of $K$. This leads to another scale, $\frac{\Lambda}{K}$ beyond which this expansion also breaks down. The origin of this scale becomes evident when both cutoffs are imposed sharply. As depicted in Fig~[\ref{VHove_Cutoffs}], the smaller $|\mathbf{q}|$ is, the farther from the origin the modes that are decimated lie. This results in nonanalyticity of $\partial_{\log \Lambda}\Gamma_{\text{\tiny PH}} (\mathbf{q},\Omega)$ as it is proportional to $\theta(1-\frac{\Lambda^2}{q^2 K^2})$. Note that this applies to both $\mu=0$ and $\mu \neq 0$ as long as $\mu\ll \Lambda,K^2$. Imposing a soft momentum cutoff, while maintaining a sharp energy cutoff, will not cure this nonanalyticity. For our momentum regulator $\exp(-|\mathbf{k}|^2/K^2)$, one finds nonanalytic dependence on $|\mathbf{q}|$ in $\partial_{\log \Lambda}\Gamma_{\text{\tiny PH}} (\mathbf{q},\Omega)$ of the form 
$\exp(-\frac{\Lambda^2}{|\mathbf{q}|^2 K^2})$. When both cutoffs are imposed softly, this aspect of the problem manifests itself as a finite convergence radius $q<\frac{\Lambda}{K}$. Unlike terms singular in $\mu$, these terms are independent of the details of the energy regulator. Since $\mu\ll\Lambda\simeq K^2$ at the initial steps of RG near the critical point, the criterion $q<\frac{\Lambda}{K}$ is automatically satisfied if $q<2\sqrt{\mu}$. Since $\mu$ and $K$ grow under this RG (or, equivalently, as the energy cutoff is lowered), there comes a point where the constraint $q<\frac{\Lambda}{K}$ becomes more stringent than $q < 2 \sqrt{\mu}$ (see Fig.~[\ref{Analytic_Window}]).

For the particle-particle diagram, which is given by Eq.~(\ref{1-loop_PP}), in the limit $\mathbf{q}=0$ and $\Omega\ll\Lambda$ we obtain: 
\begin{align}
	&\Gamma_{\text{\tiny PP}}(\mathbf{q}=0,\Omega\ll\Lambda)
	\approx
	\frac{1}{(2\pi)^2}\Bigg[
		\log\frac{K^2}{\Lambda}  
		\log\Big(1+\frac{4\Lambda^2}{\Omega^2}\Big)
	\nonumber \\
	& \qquad \qquad \qquad
		+ 
		\frac{1}{4}\,\log^2\Big(\frac{\Omega^2}{4\Lambda^2}\Big) 
		+ 
		\mathcal{O}(\frac{\Omega^2}{4\Lambda^2},\mu)
	\Bigg].
	\label{Gamma_PP_Zmu}
\end{align}
Note that the quantum effective action is nonanalytic in frequency. In particular, it contains a term that is proportional to $\log^{2}\big(\frac{\Omega^2}{4\Lambda^2}\big)$, which generates a nonlocal term in the Wilsonian effective action. However, it is cancelled by another nonlocal term generated from the first term, which is sensitive to the size of the Fermi surface. As a result, the net contribution to the Wilsonian effective action remains local:
\begin{align}
	&
	\partial_{\log\Lambda}\Gamma_{\text{\tiny PP}}(\mathbf{q}=0,\Omega\ll\Lambda) 
	\,\approx\, 
	\frac{1}{(2\pi)^2}\Bigg[
		\frac{8\Lambda^2}{4\Lambda^2+\Omega^2}\log\frac{K^2}{\Lambda}
	\nonumber \\
	& \qquad \qquad \qquad
		\,-\,\log\Big( 1 + \frac{\Omega^2}{4\Lambda^2} \Big)
		\,+\,
		\mathcal{O}(\frac{\Omega^2}{4\Lambda^2},\mu)
	\Bigg].
	\label{dLogLambda_Gamma_PP_Zmu}
\end{align}

As shown in Eq.~(\ref{Formal-beta_functions}), this contributes to the $\beta$ function for the local four-fermion interaction term. It is noted that the quantum effective action and the $\beta$ functions are ill-defined without the momentum cutoff, $K$.


\subsection{$\mu=0$} 

In the previous subsection, we found that the window of convergence for the gradient expansion of the effective action vanishes in the $\mu \rightarrow 0$ limit. This suggests that the effective action is nonanalytic at $\mu=0$. Indeed, we find that at the critical point of the neck-narrowing transition ($\mu=0$), $\partial_{\log\Lambda}\Gamma_{\text{\tiny PH}}(q\hat{x},0)$ is nonanalytic in $q$. In the limit  $q \ll \Lambda,  K$, we obtain the following series expansion in $q$:  
\begin{align}
	&\partial_{\log\Lambda}\Gamma_{\text{\tiny PH}}(q\hat{x},0)
	\approx
	\frac{1}{(2\pi)^2}
	\Big[
		-
		0.028 \frac{q^4}{\Lambda^2}
		-
		0.368 \frac{K^2}{\Lambda} \frac{q^6}{\Lambda^3}
	\nonumber 
	\\
	&
		+
		\big( 0.473\frac{K^4}{\Lambda^2} + 0.448 \big) \frac{q^8}{\Lambda^4}
		-
		\big( 0.55\frac{K^6}{\Lambda^3} + 0.6\frac{K^2}{\Lambda} \big) \frac{q^{10}}{\Lambda^5}	
	\nonumber 
	\\
	&
		+ \mathcal{O}(\frac{q^{12}}{\Lambda^6})
	\Big]
	\log\frac{q^2}{K^2}\,,
	\label{Zero_mu_dPH_qx_Kfinite}
\end{align}
where analytic terms are not shown and $\mathcal{O}(\frac{1}{K})$ terms are ignored.


\subsection{$\beta$ Functions} 

The $\beta$ functions of $g$, and the first five subleading vertices $g^{\{n\}}_{\text{\tiny PH}}\equiv g^{\{n,0\}}_{\text{\tiny PH}}$ are obtained from Eq.~(\ref{PH_NZmu_qx_l2kF}). These $\beta$ functions describe the evolution of the local effective action when $\mu>0$. From here on we set $\Lambda$ to $1$ for brevity ($\Lambda$ can be restored by $K^2\to K^2/\Lambda$ and $\mu\to \mu/\Lambda$). The one-loop $\beta$ functions are:

\begin{subequations} \label{Beta_functions}  
\begin{align}
	\dot{g} =& -\frac{g^2}{2\pi^2}\log K^2
	\label{Beta_functions:g} 
	\\[.5em]
	\dot{g}^{\{1\}}_{\text{\tiny PH}} =& -g^{\{1\}}_{\text{\tiny PH}} - \frac{g^2}{2(2\pi)^2}8 K^2
	\label{Beta_functions:g1} 
	\\[.5em]
	\dot{g}^{\{2)}_{\text{\tiny PH}} =& -2g^{\{2\}}_{\text{\tiny PH}} - \frac{g^2}{2(2\pi)^2}
	\Big( 
		19.7\log\frac{\mu}{K^2} 
		+  
		101
		-
		5.6K^4
	\Big)
	\label{Beta_functions:g2} 
	\\[.5em]
	\dot{g}^{\{3\}}_{\text{\tiny PH}} =& -3g^{\{3\}}_{\text{\tiny PH}} - \frac{g^2}{2(2\pi)^2}
	\Big( 
		-
		\frac{1.87}{\mu}
		+
		\Big[
			22.8K^2
			+
			\frac{86}{K^2}
		\Big]\log\frac{\mu}{K^2}
	\nonumber \\
	& 
		+
		4.11K^6 
		+
		162K^2 
	\Big)
	\label{Beta_functions:g3}   
	\\[.5em]
	\dot{g}^{\{4\}}_{\text{\tiny PH}} =& -4g^{\{4\}}_{\text{\tiny PH}} - \frac{g^2}{2(2\pi)^2}
	\Big( 
		-
			\frac{0.13}{\mu^2}
			-
			2.06\frac{K^2}{\mu}
			+
			\Big[
				\frac{237}{K^4}
				+
				255
	\nonumber \\
	& 
				+
				8.7 K^4
			\Big]\log\frac{\mu}{K^2}
			+
			1462
			-
			71.5K^4
			-
			3.16K^8 
	\Big)
	\label{Beta_functions:g4}  
	\\[.5em]
	\dot{g}^{\{5\}}_{\text{\tiny PH}} =& -5g^{\{5\}}_{\text{\tiny PH}} - \frac{g^2}{2(2\pi)^2}
	\Big( 
			-
			\frac{0.014}{\mu^3}
			-
			\frac{0.138 K^2}{\mu^2}
			-
			\frac{0.326}{K^2\mu^2}
	\nonumber \\
	& 
			-
			0.716\frac{K^4}{\mu}
			-
			\frac{17.13}{\mu}
			-
			\frac{12.98}{K^4\mu}
			+\Big[
				-
				1.262K^6 
	\nonumber \\
	& 
				+
				305K^2
				+
				\frac{1082}{K^2}
				+
				\frac{483}{K^6}
			\Big]\log\frac{\mu}{K^2} 
			+
			2.52K^{10}
	\nonumber \\
	& 
			+
			9.235K^6
			+
			1987K^2
	\Big)
	\label{Beta_functions:g5}
\end{align}
\end{subequations}
where, $K = K_{0} e^{\ell/2}$ and $\mu = \mu_{0} e^{\ell}$ ($\ell$ is the RG ``time'', $\mu_0 = \mu_{\ell=0}$ and $K_0 = K_{\ell=0}$ are the initial values).

\begin{figure}[h] 
        \centering
        \includegraphics[width=.45\textwidth]{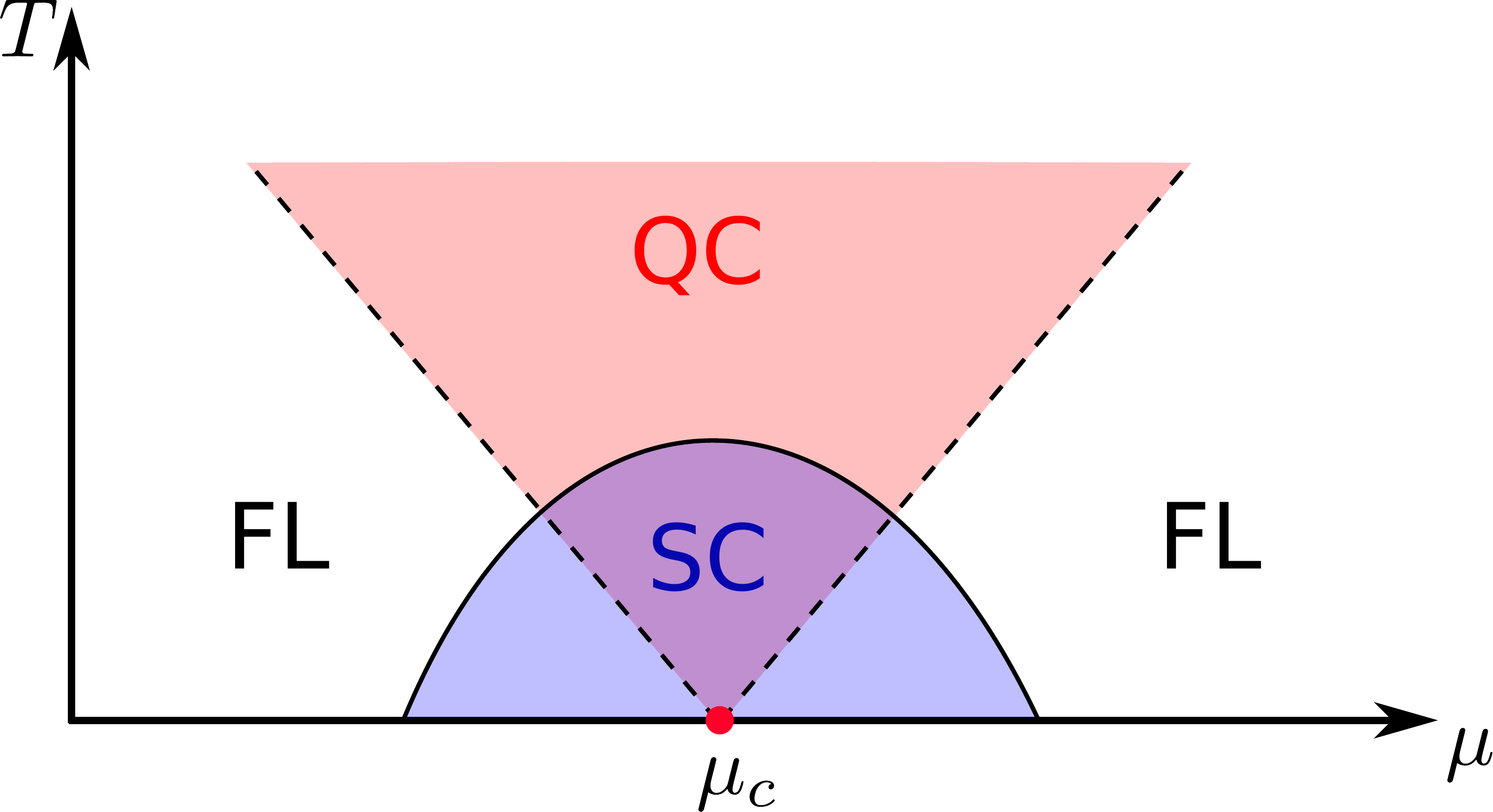}
	
        \caption[]{\justifying 
A schematic phase diagram of a neck-narrowing Lifshitz transition in the presence of weak attractive interactions. 
In the quantum critical (QC) region, the pairing susceptibility grows as $\log^{2} T$ in temperature $T$.
Away from the critical point, the system enters into the Fermi liquid (FL) region below the temperature scale $T \sim |\mu - \mu_c|$ where the pairing susceptibility grows as $-\log T$.
However, it is likely that the system undergoes a superconducting (SC) phase transition before it crosses over into the Fermi liquid regime near the critical point which is indicated by the dome in the figure.
Even the Fermi liquid will eventually become unstable toward SC state at sufficiently low temperatures, which is not indicated in the figure.
}
	\label{Cartoon_Phase_Diagram}
\end{figure}

The $\beta$ function of $g$ can be solved analytically:
\begin{equation}
	g(\ell) \,=\, \frac{2\pi^2}{\ell\log K_0^2+\frac{1}{2}\ell^2+\frac{2\pi^2}{g_0}}\,,
	\label{g_ell}
\end{equation}
where $g_0 = g_{\ell=0}$. Thus, $g$ grows (decreases) at low energies for attractive (repulsive) interaction. It is noted that $g$ runs quadratically in the logarithmic length scale $\ell$. One of the logarithms originates from the usual BCS enhancement of an attractive interaction. 
The other logarithm reflects the fact that the size of the Fermi surface, $K$, increases under the scaling that expands the momentum space with respect to ${\bf k}=0$. Of course, we have to consider the fact that $\mu$ also grows under this scaling. Once $\mu$ becomes comparable with $\Lambda$, the van Hove singular point is no longer a special point and we should use the alternative scheme (\textit{{\`a} la} Shankar) where momenta are scaled towards the nearest points on the Fermi surface. This implies that the attractive interaction increases quadratically in $\ell$ for $\ell < \ell_s \sim \log \Lambda/\mu$. The quartic coupling reaches order of unity at the scale  $\ell^{*}\,=\, -\log K_0^2+\sqrt{\log^2(K_0^2)+\frac{2}{|g_0|}+4\pi^2}$, 
which is smaller than $\ell_s$ for $\mu \ll \Lambda$. 
This suggests that, when the system is sufficiently close to the van Hove singularity, it can become unstable toward a superconducting state before entering into a regime controlled by the usual Fermi liquid that exhibits a simple logarithmic growth of the pairing interaction. 
A schematic phase diagram is illustrated in Fig.~[\ref{Cartoon_Phase_Diagram}].
It is emphasized that we are able to capture the superlogarithmic growth of the coupling within a local RG scheme owing to the introduction of the momentum cutoff $K$, which is treated as a dimensionful coupling.
The rest of the $\beta$ functions can be solved numerically.  

\begin{figure}
        \centering
        \includegraphics[width=0.4\textwidth]{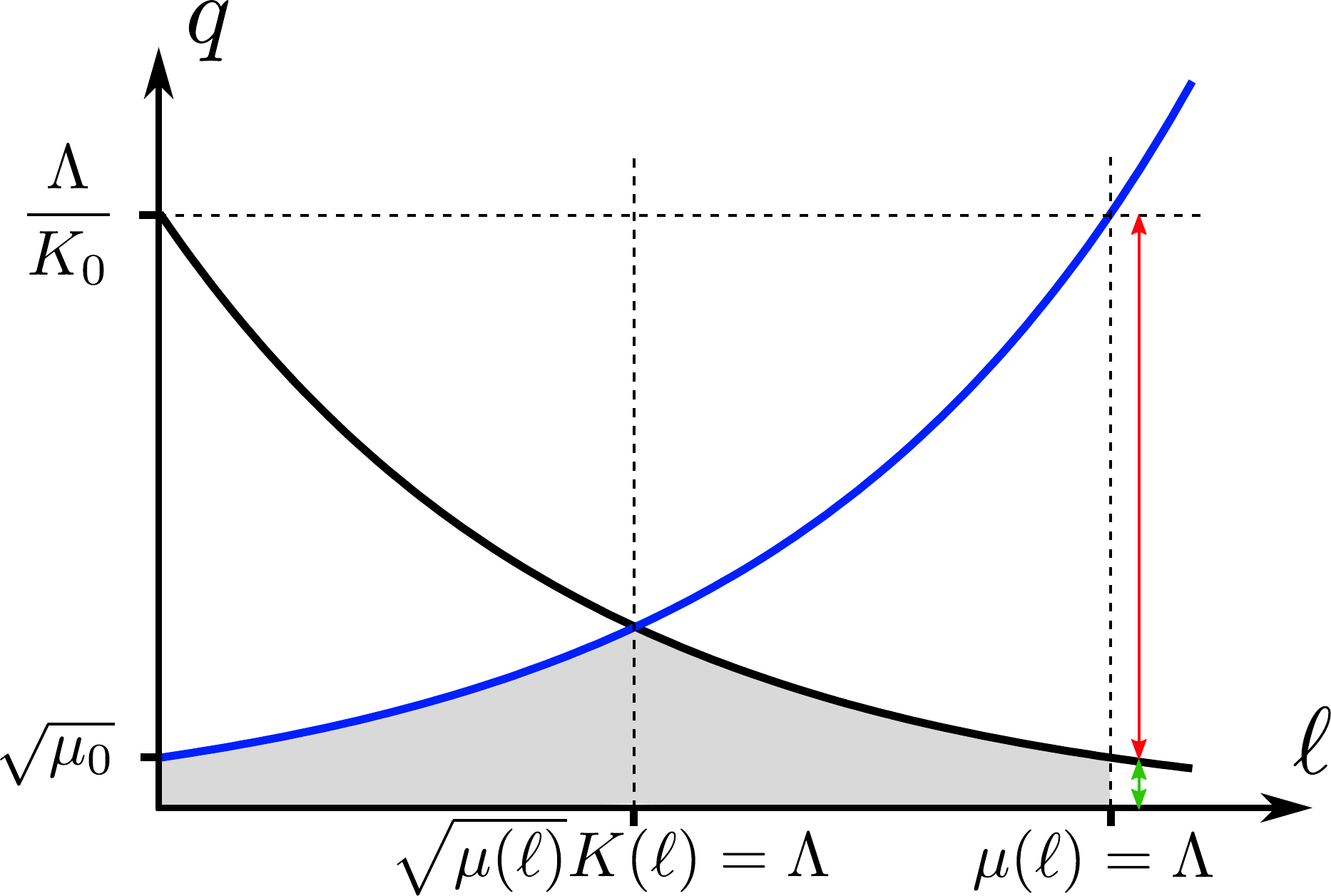}
         
        \caption[]{\justifying Depiction of the change of the analytic window (the shaded region below the two curves) in the course of RG. 
Initially, the width of the neck (due to the $2k_F$ physics) determines this analytic window (blue curve). 
Beyond the point $\sqrt{\mu(\ell)} K(\ell)=\Lambda$, it is $\Lambda/K$ that dictates analytic range of quantum corrections (the black curve). 
Starting from a microscopic scale ($\Lambda\simeq K^2$), by the time the energy cutoff has been lowered to $\mu$, the analytic window has shrunk to $\mathcal{O}(\frac{\mu}{K})$.}
	\label{Analytic_Window}
\end{figure}

Observe that, the $\beta$ functions of subleading couplings contain higher powers of $K$. 
For example, there is a $K^{10}$ term in the $\beta$ function of $g^{\{5\}}_{\text{\tiny PH}}$ in Eq.~(\ref{Beta_functions:g5}). This is the reflection of the fact that the effective action becomes nonanalytic in the large $K$ limit. The analytic window, within which one can describe the evolution of effective interactions by the above $\beta$ functions, changes in the course of RG as shown in Fig.~[\ref{Analytic_Window}]. Regardless of the details of the cutoffs (even for a sharp cutoff), by the time the energy cutoff has been sufficiently lowered ($\Lambda\lesssim\mu$) this analytic window has shrunk to $\mathcal{O}(\frac{\mu}{K})$. 
This indicates the strong dependence of quantum corrections on $\mathbf{q}$ that cannot be captured entirely in terms of local operators.

%

\section{Summary and Discussion}
\label{conclusions}

In summary, we have demonstrated that the locality of the effective action can be retained only away from the Lifshitz critical point ($\mu\neq0$) and in the presence of a momentum cutoff. Based on a local renormalization group scheme implemented near the critical point, we show that a short-range attractive interaction grows as $\log^2 E$ at energy scales above the chemical potential. The fast growth of the pairing interaction, which is distinct from the simple logarithmic growth in the low-energy Fermi liquid regime ($E<\mu$), implies that near the critical point the system is likely to undergo a superconducting phase transition before it enters into the Fermi liquid.

The Wilsonian effective action is analytic within a finite momentum range, which shrinks as the critical point is approached. This indicates that the Wilsonian effective action is intrinsically nonlocal in the presence of interactions at the Lifshitz quantum critical point in two space dimensions. The intrinsic nonlocality in the Wilsonian effective action right at the critical point suggests that some gapless modes have been integrated out in the coarse-graining procedure. To understand what is required to keep the locality at the critical point, it is illuminating to compare the case at hand with quantum critical points associated with more conventional symmetry-breaking phase transitions in metals. If one insists on keeping only electron fields to describe a symmetry-breaking quantum critical point at low energies, one inevitably encounters a nonlocality in the Wilsonian effective action. This is because gapless order parameter fluctuations are ``integrated'' out in the pure fermionic description. In order to keep locality, one has to explicitly include a gapless collective mode for the order parameter. Analogously, we envisage the emergence of a gapless mode at the Lifshitz critical point, associated with the critical fluctuations of the Fermi surface topology. Unlike in symmetry-breaking phase transitions, the gapless mode in the Lifshitz transition should describe global degrees of freedom. In the future, it will be of great interest to find a local description for the critical point, by including an extra mode that becomes gapless at the critical point.

%
%
\begin{acknowledgements}

The authors would like to thank Anton Kapustin, Steven Kivelson and Subhro Bhattacharjee for useful discussions. 
This work is supported in part by the Natural Sciences and Engineering Research Council of Canada (S.S.L. and C.K.), the Early Research Award from the Ontario Ministry of Research and Innovation and the Templeton Foundation (S.S.L.), CIFAR at McMaster, and by the Canada Research Chair and Canada Council Killam programs and the National Science Foundation under Grant No. NSF PHY11-25915 (C.K.). 
Research at the Perimeter Institute is supported in part by the Government of Canada through Industry Canada, and by the Province of Ontario through the Ministry of Research and Information. 

\end{acknowledgements}
%
%
%
%
%
\end{fmffile} 
\bibliographystyle{apsrev4-1}
\bibliography{./VHove_Modified}

\begin{thebibliography}{30}%
\makeatletter
\providecommand \@ifxundefined [1]{%
 \@ifx{#1\undefined}
}%
\providecommand \@ifnum [1]{%
 \ifnum #1\expandafter \@firstoftwo
 \else \expandafter \@secondoftwo
 \fi
}%
\providecommand \@ifx [1]{%
 \ifx #1\expandafter \@firstoftwo
 \else \expandafter \@secondoftwo
 \fi
}%
\providecommand \natexlab [1]{#1}%
\providecommand \enquote  [1]{``#1''}%
\providecommand \bibnamefont  [1]{#1}%
\providecommand \bibfnamefont [1]{#1}%
\providecommand \citenamefont [1]{#1}%
\providecommand \href@noop [0]{\@secondoftwo}%
\providecommand \href [0]{\begingroup \@sanitize@url \@href}%
\providecommand \@href[1]{\@@startlink{#1}\@@href}%
\providecommand \@@href[1]{\endgroup#1\@@endlink}%
\providecommand \@sanitize@url [0]{\catcode `\\12\catcode `\$12\catcode
  `\&12\catcode `\#12\catcode `\^12\catcode `\_12\catcode `\%12\relax}%
\providecommand \@@startlink[1]{}%
\providecommand \@@endlink[0]{}%
\providecommand \url  [0]{\begingroup\@sanitize@url \@url }%
\providecommand \@url [1]{\endgroup\@href {#1}{\urlprefix }}%
\providecommand \urlprefix  [0]{URL }%
\providecommand \Eprint [0]{\href }%
\providecommand \doibase [0]{http://dx.doi.org/}%
\providecommand \selectlanguage [0]{\@gobble}%
\providecommand \bibinfo  [0]{\@secondoftwo}%
\providecommand \bibfield  [0]{\@secondoftwo}%
\providecommand \translation [1]{[#1]}%
\providecommand \BibitemOpen [0]{}%
\providecommand \bibitemStop [0]{}%
\providecommand \bibitemNoStop [0]{.\EOS\space}%
\providecommand \EOS [0]{\spacefactor3000\relax}%
\providecommand \BibitemShut  [1]{\csname bibitem#1\endcsname}%
\let\auto@bib@innerbib\@empty
\bibitem [{\citenamefont {Lifshitz}(1960)}]{Lifshitz1960}%
  \BibitemOpen
  \bibfield  {author} {\bibinfo {author} {\bibfnamefont {I.~M.}\ \bibnamefont
  {Lifshitz}},\ }\href@noop {} {\bibfield  {journal} {\bibinfo  {journal} {Sov.
  Phys. JETP}\ }\textbf {\bibinfo {volume} {11}},\ \bibinfo {pages} {1130}
  (\bibinfo {year} {1960})}\BibitemShut {NoStop}%
\bibitem [{\citenamefont {Yamaji}\ \emph {et~al.}(2006)\citenamefont {Yamaji},
  \citenamefont {Misawa},\ and\ \citenamefont {Imada}}]{Yamaji2006}%
  \BibitemOpen
  \bibfield  {author} {\bibinfo {author} {\bibfnamefont {Y.}~\bibnamefont
  {Yamaji}}, \bibinfo {author} {\bibfnamefont {T.}~\bibnamefont {Misawa}}, \
  and\ \bibinfo {author} {\bibfnamefont {M.}~\bibnamefont {Imada}},\ }\href
  {\doibase 10.1143/JPSJ.75.094719} {\bibfield  {journal} {\bibinfo  {journal}
  {J. Phys. Soc. Jpn.}\ }\textbf {\bibinfo {volume} {75}},\ \bibinfo {pages}
  {094719} (\bibinfo {year} {2006})}\BibitemShut {NoStop}%
\bibitem [{\citenamefont {Wen}(2004)}]{WEN2004}%
  \BibitemOpen
  \bibfield  {author} {\bibinfo {author} {\bibfnamefont {X.-G.}\ \bibnamefont
  {Wen}},\ }\href {http://ukcatalogue.oup.com/product/9780198530947.do} {\emph
  {\bibinfo {title} {Quantum Field Theory of Many-body Systems}}},\ Oxford
  Graduate Texts\ (\bibinfo  {publisher} {Oxford University Press, New York},\
  \bibinfo {year} {2004})\BibitemShut {NoStop}%
\bibitem [{\citenamefont {Rodney}\ \emph {et~al.}(2013)\citenamefont {Rodney},
  \citenamefont {Song}, \citenamefont {Lee}, \citenamefont {{Le Hur}},\ and\
  \citenamefont {Sorensen}}]{Rodney2013}%
  \BibitemOpen
  \bibfield  {author} {\bibinfo {author} {\bibfnamefont {M.}~\bibnamefont
  {Rodney}}, \bibinfo {author} {\bibfnamefont {H.~F.}\ \bibnamefont {Song}},
  \bibinfo {author} {\bibfnamefont {S.-S.}\ \bibnamefont {Lee}}, \bibinfo
  {author} {\bibfnamefont {K.}~\bibnamefont {{Le Hur}}}, \ and\ \bibinfo
  {author} {\bibfnamefont {E.~S.}\ \bibnamefont {Sorensen}},\ }\href {\doibase
  10.1103/PhysRevB.87.115132} {\bibfield  {journal} {\bibinfo  {journal} {Phys.
  Rev. B}\ }\textbf {\bibinfo {volume} {87}},\ \bibinfo {pages} {115132}
  (\bibinfo {year} {2013})}\BibitemShut {NoStop}%
\bibitem [{\citenamefont {Schulz}(1987)}]{Schulz1987}%
  \BibitemOpen
  \bibfield  {author} {\bibinfo {author} {\bibfnamefont {H.~J.}\ \bibnamefont
  {Schulz}},\ }\href {\doibase 10.1209/0295-5075/4/5/016} {\bibfield  {journal}
  {\bibinfo  {journal} {Europhys. Lett.}\ }\textbf {\bibinfo {volume} {4}},\
  \bibinfo {pages} {609} (\bibinfo {year} {1987})}\BibitemShut {NoStop}%
\bibitem [{\citenamefont {Lederer}\ \emph {et~al.}(1987)\citenamefont
  {Lederer}, \citenamefont {Montambaux},\ and\ \citenamefont
  {Poilblanc}}]{Lederer1987}%
  \BibitemOpen
  \bibfield  {author} {\bibinfo {author} {\bibfnamefont {P.}~\bibnamefont
  {Lederer}}, \bibinfo {author} {\bibfnamefont {G.}~\bibnamefont {Montambaux}},
  \ and\ \bibinfo {author} {\bibfnamefont {D.}~\bibnamefont {Poilblanc}},\
  }\href {http://www.edpsciences.org/10.1051/jphys:0198700480100161300}
  {\bibfield  {journal} {\bibinfo  {journal} {J. Phys. (Paris)}\ }\textbf
  {\bibinfo {volume} {48}},\ \bibinfo {pages} {1613} (\bibinfo {year}
  {1987})}\BibitemShut {NoStop}%
\bibitem [{\citenamefont {Gonz\'{a}lez}\ \emph {et~al.}(1997)\citenamefont
  {Gonz\'{a}lez}, \citenamefont {Guinea},\ and\ \citenamefont
  {Vozmediano}}]{Gonzalez1997}%
  \BibitemOpen
  \bibfield  {author} {\bibinfo {author} {\bibfnamefont {J.}~\bibnamefont
  {Gonz\'{a}lez}}, \bibinfo {author} {\bibfnamefont {F.}~\bibnamefont
  {Guinea}}, \ and\ \bibinfo {author} {\bibfnamefont {M.~M. A.~H.}\
  \bibnamefont {Vozmediano}},\ }\href
  {http://linkinghub.elsevier.com/retrieve/pii/S0550321396006207} {\bibfield
  {journal} {\bibinfo  {journal} {Nucl. Phys. B}\ }\textbf {\bibinfo {volume}
  {485}},\ \bibinfo {pages} {694} (\bibinfo {year} {1997})}\BibitemShut
  {NoStop}%
\bibitem [{\citenamefont {Alvarez}\ \emph {et~al.}(1998)\citenamefont
  {Alvarez}, \citenamefont {Gonz\'{a}lez}, \citenamefont {Guinea},\ and\
  \citenamefont {Vozmediano}}]{Alvarez1998}%
  \BibitemOpen
  \bibfield  {author} {\bibinfo {author} {\bibfnamefont {J.~V.}\ \bibnamefont
  {Alvarez}}, \bibinfo {author} {\bibfnamefont {J.}~\bibnamefont
  {Gonz\'{a}lez}}, \bibinfo {author} {\bibfnamefont {F.}~\bibnamefont
  {Guinea}}, \ and\ \bibinfo {author} {\bibfnamefont {M.~A.~H.}\ \bibnamefont
  {Vozmediano}},\ }\href {\doibase 10.1143/JPSJ.67.1868} {\bibfield  {journal}
  {\bibinfo  {journal} {J. Phys. Soc. Jpn.}\ }\textbf {\bibinfo {volume}
  {67}},\ \bibinfo {pages} {1868} (\bibinfo {year} {1998})}\BibitemShut
  {NoStop}%
\bibitem [{\citenamefont {Furukawa}\ \emph {et~al.}(1998)\citenamefont
  {Furukawa}, \citenamefont {Rice},\ and\ \citenamefont
  {Salmhofer}}]{Furukawa1998}%
  \BibitemOpen
  \bibfield  {author} {\bibinfo {author} {\bibfnamefont {N.}~\bibnamefont
  {Furukawa}}, \bibinfo {author} {\bibfnamefont {T.~M.}\ \bibnamefont {Rice}},
  \ and\ \bibinfo {author} {\bibfnamefont {M.}~\bibnamefont {Salmhofer}},\
  }\href {http://link.aps.org/doi/10.1103/PhysRevLett.81.3195} {\bibfield
  {journal} {\bibinfo  {journal} {Phys. Rev. Lett.}\ }\textbf {\bibinfo
  {volume} {81}},\ \bibinfo {pages} {3195} (\bibinfo {year}
  {1998})}\BibitemShut {NoStop}%
\bibitem [{\citenamefont {Furukawa}\ \emph {et~al.}(2000)\citenamefont
  {Furukawa}, \citenamefont {Honerkamp}, \citenamefont {Salmhofer},\ and\
  \citenamefont {Rice}}]{Furukawa2000}%
  \BibitemOpen
  \bibfield  {author} {\bibinfo {author} {\bibfnamefont {N.}~\bibnamefont
  {Furukawa}}, \bibinfo {author} {\bibfnamefont {C.}~\bibnamefont {Honerkamp}},
  \bibinfo {author} {\bibfnamefont {M.}~\bibnamefont {Salmhofer}}, \ and\
  \bibinfo {author} {\bibfnamefont {T.~M.}\ \bibnamefont {Rice}},\ }\href
  {\doibase 10.1016/S0921-4526(99)02824-0} {\bibfield  {journal} {\bibinfo
  {journal} {Phys. B (Amsterdam, Neth.)}\ }\textbf {\bibinfo {volume} {284}},\
  \bibinfo {pages} {1571} (\bibinfo {year} {2000})}\BibitemShut {NoStop}%
\bibitem [{\citenamefont {Gonz\'{a}lez}(2001)}]{Gonzalez2001}%
  \BibitemOpen
  \bibfield  {author} {\bibinfo {author} {\bibfnamefont {J.}~\bibnamefont
  {Gonz\'{a}lez}},\ }\href {\doibase 10.1103/PhysRevB.63.045114} {\bibfield
  {journal} {\bibinfo  {journal} {Phys. Rev. B}\ }\textbf {\bibinfo {volume}
  {63}},\ \bibinfo {pages} {1} (\bibinfo {year} {2001})}\BibitemShut {NoStop}%
\bibitem [{\citenamefont {{Le Hur}}\ and\ \citenamefont
  {Rice}(2009)}]{LeHur2009}%
  \BibitemOpen
  \bibfield  {author} {\bibinfo {author} {\bibfnamefont {K.}~\bibnamefont {{Le
  Hur}}}\ and\ \bibinfo {author} {\bibfnamefont {T.~M.}\ \bibnamefont {Rice}},\
  }\href {http://dx.doi.org/10.1016/j.aop.2009.02.004} {\bibfield  {journal}
  {\bibinfo  {journal} {Ann. Phys. (N.Y.)}\ }\textbf {\bibinfo {volume}
  {324}},\ \bibinfo {pages} {1452} (\bibinfo {year} {2009})}\BibitemShut
  {NoStop}%
\bibitem [{\citenamefont {Nandkishore}\ \emph {et~al.}(2012)\citenamefont
  {Nandkishore}, \citenamefont {Levitov},\ and\ \citenamefont
  {Chubukov}}]{Nandkishore2012}%
  \BibitemOpen
  \bibfield  {author} {\bibinfo {author} {\bibfnamefont {R.}~\bibnamefont
  {Nandkishore}}, \bibinfo {author} {\bibfnamefont {L.~S.}\ \bibnamefont
  {Levitov}}, \ and\ \bibinfo {author} {\bibfnamefont {A.~V.}\ \bibnamefont
  {Chubukov}},\ }\href {http://www.nature.com/doifinder/10.1038/nphys2208}
  {\bibfield  {journal} {\bibinfo  {journal} {Nature Physics}\ }\textbf
  {\bibinfo {volume} {8}},\ \bibinfo {pages} {158} (\bibinfo {year}
  {2012})}\BibitemShut {NoStop}%
\bibitem [{\citenamefont {Kapustin}\ and\ \citenamefont
  {Rothstein}()}]{Kapustin2012}%
  \BibitemOpen
  \bibfield  {author} {\bibinfo {author} {\bibfnamefont {A.}~\bibnamefont
  {Kapustin}}\ and\ \bibinfo {author} {\bibfnamefont {I.}~\bibnamefont
  {Rothstein}},\ }\href@noop {} {}\bibinfo {note} {(unpublished)}\BibitemShut
  {NoStop}%
\bibitem [{\citenamefont {Gonz\'alez}(2013)}]{Gonzalez2013}%
  \BibitemOpen
  \bibfield  {author} {\bibinfo {author} {\bibfnamefont {J.}~\bibnamefont
  {Gonz\'alez}},\ }\href {\doibase 10.1103/PhysRevB.88.125434} {\bibfield
  {journal} {\bibinfo  {journal} {Phys. Rev. B}\ }\textbf {\bibinfo {volume}
  {88}},\ \bibinfo {pages} {125434} (\bibinfo {year} {2013})}\BibitemShut
  {NoStop}%
\bibitem [{\citenamefont {Yudin}\ \emph {et~al.}(2014)\citenamefont {Yudin},
  \citenamefont {Hirschmeier}, \citenamefont {Hafermann}, \citenamefont
  {Eriksson}, \citenamefont {Lichtenstein},\ and\ \citenamefont
  {Katsnelson}}]{Yudin2013}%
  \BibitemOpen
  \bibfield  {author} {\bibinfo {author} {\bibfnamefont {D.}~\bibnamefont
  {Yudin}}, \bibinfo {author} {\bibfnamefont {D.}~\bibnamefont {Hirschmeier}},
  \bibinfo {author} {\bibfnamefont {H.}~\bibnamefont {Hafermann}}, \bibinfo
  {author} {\bibfnamefont {O.}~\bibnamefont {Eriksson}}, \bibinfo {author}
  {\bibfnamefont {A.~I.}\ \bibnamefont {Lichtenstein}}, \ and\ \bibinfo
  {author} {\bibfnamefont {M.~I.}\ \bibnamefont {Katsnelson}},\ }\href
  {\doibase 10.1103/PhysRevLett.112.070403} {\bibfield  {journal} {\bibinfo
  {journal} {Phys. Rev. Lett.}\ }\textbf {\bibinfo {volume} {112}},\ \bibinfo
  {pages} {070403} (\bibinfo {year} {2014})}\BibitemShut {NoStop}%
\bibitem [{\citenamefont {Khavkine}\ \emph {et~al.}(2004)\citenamefont
  {Khavkine}, \citenamefont {Chung}, \citenamefont {Oganesyan},\ and\
  \citenamefont {Kee}}]{Khavkine2004}%
  \BibitemOpen
  \bibfield  {author} {\bibinfo {author} {\bibfnamefont {I.}~\bibnamefont
  {Khavkine}}, \bibinfo {author} {\bibfnamefont {C.-H.}\ \bibnamefont {Chung}},
  \bibinfo {author} {\bibfnamefont {V.}~\bibnamefont {Oganesyan}}, \ and\
  \bibinfo {author} {\bibfnamefont {H.-Y.}\ \bibnamefont {Kee}},\ }\href
  {\doibase 10.1103/PhysRevB.70.155110} {\bibfield  {journal} {\bibinfo
  {journal} {Phys. Rev. B}\ }\textbf {\bibinfo {volume} {70}},\ \bibinfo
  {pages} {1} (\bibinfo {year} {2004})}\BibitemShut {NoStop}%
\bibitem [{\citenamefont {Yamase}\ \emph {et~al.}(2005)\citenamefont {Yamase},
  \citenamefont {Oganesyan},\ and\ \citenamefont {Metzner}}]{Yamase2005}%
  \BibitemOpen
  \bibfield  {author} {\bibinfo {author} {\bibfnamefont {H.}~\bibnamefont
  {Yamase}}, \bibinfo {author} {\bibfnamefont {V.}~\bibnamefont {Oganesyan}}, \
  and\ \bibinfo {author} {\bibfnamefont {W.}~\bibnamefont {Metzner}},\ }\href
  {\doibase 10.1103/PhysRevB.72.035114} {\bibfield  {journal} {\bibinfo
  {journal} {Phys. Rev. B}\ }\textbf {\bibinfo {volume} {72}},\ \bibinfo
  {pages} {035114} (\bibinfo {year} {2005})}\BibitemShut {NoStop}%
\bibitem [{\citenamefont {Zanchi}\ and\ \citenamefont
  {Schulz}(1996)}]{Zanchi1996}%
  \BibitemOpen
  \bibfield  {author} {\bibinfo {author} {\bibfnamefont {D.}~\bibnamefont
  {Zanchi}}\ and\ \bibinfo {author} {\bibfnamefont {H.}~\bibnamefont
  {Schulz}},\ }\href {http://www.ncbi.nlm.nih.gov/pubmed/9984691} {\bibfield
  {journal} {\bibinfo  {journal} {Phys. Rev. B}\ }\textbf {\bibinfo {volume}
  {54}},\ \bibinfo {pages} {9509} (\bibinfo {year} {1996})}\BibitemShut
  {NoStop}%
\bibitem [{\citenamefont {Zanchi}\ and\ \citenamefont
  {Schulz}(2000)}]{Zanchi2000}%
  \BibitemOpen
  \bibfield  {author} {\bibinfo {author} {\bibfnamefont {D.}~\bibnamefont
  {Zanchi}}\ and\ \bibinfo {author} {\bibfnamefont {H.}~\bibnamefont
  {Schulz}},\ }\href {\doibase 10.1103/PhysRevB.61.13609} {\bibfield  {journal}
  {\bibinfo  {journal} {Phys. Rev. B}\ }\textbf {\bibinfo {volume} {61}},\
  \bibinfo {pages} {13609} (\bibinfo {year} {2000})}\BibitemShut {NoStop}%
\bibitem [{\citenamefont {Reiss}\ \emph {et~al.}(2007)\citenamefont {Reiss},
  \citenamefont {Rohe},\ and\ \citenamefont {Metzner}}]{Reiss2007}%
  \BibitemOpen
  \bibfield  {author} {\bibinfo {author} {\bibfnamefont {J.}~\bibnamefont
  {Reiss}}, \bibinfo {author} {\bibfnamefont {D.}~\bibnamefont {Rohe}}, \ and\
  \bibinfo {author} {\bibfnamefont {W.}~\bibnamefont {Metzner}},\ }\href
  {\doibase 10.1103/PhysRevB.75.075110} {\bibfield  {journal} {\bibinfo
  {journal} {Phys. Rev. B}\ }\textbf {\bibinfo {volume} {75}},\ \bibinfo
  {pages} {075110} (\bibinfo {year} {2007})}\BibitemShut {NoStop}%
\bibitem [{\citenamefont {Honerkamp}\ \emph {et~al.}(2002)\citenamefont
  {Honerkamp}, \citenamefont {Salmhofer},\ and\ \citenamefont
  {Rice}}]{Honerkamp2002}%
  \BibitemOpen
  \bibfield  {author} {\bibinfo {author} {\bibfnamefont {C.}~\bibnamefont
  {Honerkamp}}, \bibinfo {author} {\bibfnamefont {M.}~\bibnamefont
  {Salmhofer}}, \ and\ \bibinfo {author} {\bibfnamefont {T.}~\bibnamefont
  {Rice}},\ }\href {\doibase 10.1140/epjb/e20020137} {\bibfield  {journal}
  {\bibinfo  {journal} {Eur. Phys. J. B}\ }\textbf {\bibinfo {volume} {27}},\
  \bibinfo {pages} {127} (\bibinfo {year} {2002})}\BibitemShut {NoStop}%
\bibitem [{\citenamefont {Neumayr}\ and\ \citenamefont
  {Metzner}(2003)}]{Neumayr2003}%
  \BibitemOpen
  \bibfield  {author} {\bibinfo {author} {\bibfnamefont {A.}~\bibnamefont
  {Neumayr}}\ and\ \bibinfo {author} {\bibfnamefont {W.}~\bibnamefont
  {Metzner}},\ }\href {\doibase 10.1103/PhysRevB.67.035112} {\bibfield
  {journal} {\bibinfo  {journal} {Phys. Rev. B}\ }\textbf {\bibinfo {volume}
  {67}},\ \bibinfo {pages} {035112} (\bibinfo {year} {2003})}\BibitemShut
  {NoStop}%
\bibitem [{\citenamefont {Raghu}\ \emph {et~al.}(2010)\citenamefont {Raghu},
  \citenamefont {Kivelson},\ and\ \citenamefont {Scalapino}}]{Raghu2010}%
  \BibitemOpen
  \bibfield  {author} {\bibinfo {author} {\bibfnamefont {S.}~\bibnamefont
  {Raghu}}, \bibinfo {author} {\bibfnamefont {S.~A.}\ \bibnamefont {Kivelson}},
  \ and\ \bibinfo {author} {\bibfnamefont {D.~J.}\ \bibnamefont {Scalapino}},\
  }\href {\doibase 10.1103/PhysRevB.81.224505} {\bibfield  {journal} {\bibinfo
  {journal} {Phys. Rev. B}\ }\textbf {\bibinfo {volume} {81}},\ \bibinfo
  {pages} {224505} (\bibinfo {year} {2010})}\BibitemShut {NoStop}%
\bibitem [{\citenamefont {Gonz\'{a}lez}\ \emph {et~al.}(2000)\citenamefont
  {Gonz\'{a}lez}, \citenamefont {Guinea},\ and\ \citenamefont
  {Vozmediano}}]{Gonzalez2000}%
  \BibitemOpen
  \bibfield  {author} {\bibinfo {author} {\bibfnamefont {F.}~\bibnamefont
  {Gonz\'{a}lez}}, \bibinfo {author} {\bibfnamefont {F.}~\bibnamefont
  {Guinea}}, \ and\ \bibinfo {author} {\bibfnamefont {M.~A.~H.}\ \bibnamefont
  {Vozmediano}},\ }\href
  {http://journals.aps.org/prl/abstract/10.1103/PhysRevLett.84.4930} {\bibfield
   {journal} {\bibinfo  {journal} {Phys. Rev. Lett.}\ }\textbf {\bibinfo
  {volume} {84}},\ \bibinfo {pages} {4930} (\bibinfo {year}
  {2000})}\BibitemShut {NoStop}%
\bibitem [{\citenamefont {{Mandal}}\ and\ \citenamefont
  {{Lee}}(2015)}]{Mandal2015}%
  \BibitemOpen
  \bibfield  {author} {\bibinfo {author} {\bibfnamefont {I.}~\bibnamefont
  {{Mandal}}}\ and\ \bibinfo {author} {\bibfnamefont {S.-S.}\ \bibnamefont
  {{Lee}}},\ }\href
  {http://journals.aps.org/prb/abstract/10.1103/PhysRevB.92.035141} {\bibfield
  {journal} {\bibinfo  {journal} {Phys. Rev. B}\ }\textbf {\bibinfo {volume}
  {92}},\ \bibinfo {pages} {035141} (\bibinfo {year} {2015})}\BibitemShut
  {NoStop}%
\bibitem [{\citenamefont {Son}(1999)}]{Son1999}%
  \BibitemOpen
  \bibfield  {author} {\bibinfo {author} {\bibfnamefont {D.}~\bibnamefont
  {Son}},\ }\href {\doibase 10.1103/PhysRevD.59.094019} {\bibfield  {journal}
  {\bibinfo  {journal} {Phys. Rev. D}\ }\textbf {\bibinfo {volume} {59}},\
  \bibinfo {pages} {094019} (\bibinfo {year} {1999})}\BibitemShut {NoStop}%
\bibitem [{\citenamefont {Metlitski}\ \emph {et~al.}(2015)\citenamefont
  {Metlitski}, \citenamefont {Mross}, \citenamefont {Sachdev},\ and\
  \citenamefont {Senthil}}]{Metlitski2014}%
  \BibitemOpen
  \bibfield  {author} {\bibinfo {author} {\bibfnamefont {M.~A.}\ \bibnamefont
  {Metlitski}}, \bibinfo {author} {\bibfnamefont {D.~F.}\ \bibnamefont
  {Mross}}, \bibinfo {author} {\bibfnamefont {S.}~\bibnamefont {Sachdev}}, \
  and\ \bibinfo {author} {\bibfnamefont {T.}~\bibnamefont {Senthil}},\ }\href
  {http://journals.aps.org/prb/abstract/10.1103/PhysRevB.91.115111} {\bibfield
  {journal} {\bibinfo  {journal} {Phys. Rev. B}\ }\textbf {\bibinfo {volume}
  {91}},\ \bibinfo {pages} {115111} (\bibinfo {year} {2015})}\BibitemShut
  {NoStop}%
\bibitem [{\citenamefont {Shankar}(1994)}]{Shankar1994}%
  \BibitemOpen
  \bibfield  {author} {\bibinfo {author} {\bibfnamefont {R.}~\bibnamefont
  {Shankar}},\ }\href@noop {} {\bibfield  {journal} {\bibinfo  {journal} {Rev.
  Mod. Phys.}\ }\textbf {\bibinfo {volume} {66}},\ \bibinfo {pages} {129}
  (\bibinfo {year} {1994})}\BibitemShut {NoStop}%
\bibitem [{\citenamefont {Polchinski}(1992)}]{Polchinski1992}%
  \BibitemOpen
  \bibfield  {author} {\bibinfo {author} {\bibfnamefont {J.}~\bibnamefont
  {Polchinski}},\ }\href {http://arxiv.org/abs/hep-th/9210046} {\enquote
  {\bibinfo {title} {{Effective Field Theory and the Fermi Surface}},}\ }
  (\bibinfo {year} {1992}),\ \Eprint {http://arxiv.org/abs/hep-ph/9210.046}
  {arXiv:hep-ph/9210.046} \BibitemShut {NoStop}%
\end{thebibliography}%
\end{document}